%% file: B2G-17-013_temp.tex
\pdfoutput=1

\documentclass[11pt,twoside,a4paper,cmspaper,final,collab]{cms-tdr}
\def\svnVersion{475826P}\def\cmsCernNoTag{CERN-EP-2018-037}\def\cmsCernDate{\today}\def\cmsMessage{Published in the Journal of High Energy Physics as \href{http://dx.doi.org/10.1007/JHEP09(2018)101}{\doi{10.1007/JHEP09(2018)101.}}}

\begin{document}\cmsNoteHeader{B2G-17-013}

\hyphenation{had-ron-i-za-tion}
\hyphenation{cal-or-i-me-ter}
\hyphenation{de-vices}
\RCS$HeadURL: svn+ssh://svn.cern.ch/reps/tdr2/papers/B2G-17-013/trunk/B2G-17-013.tex $
\RCS$Id: B2G-17-013.tex 475823 2018-09-22 14:00:46Z jpazzini $

\newlength\cmsFigWidth
\setlength\cmsFigWidth{0.425\textwidth}
\providecommand{\cmsLeft}{left\xspace}
\providecommand{\cmsRight}{right\xspace}

\newcommand{\X}{\ensuremath{\cmsSymbolFace{X}}\xspace}
\newcommand{\V}{\ensuremath{\cmsSymbolFace{V}}\xspace}
\newcommand{\Ztoll}{\ensuremath{\PZ\to\ell\ell}\xspace}
\newcommand{\Ztott}{\ensuremath{\PZ\to\tau\tau}\xspace}
\newcommand{\llqq}{\ensuremath{2\ell2\cPq}\xspace}
\newcommand{\Top}{\ensuremath{\text{Top}}\xspace}
\newcommand{\VV}{\ensuremath{\V\V}\xspace}
\newcommand{\VZ}{\ensuremath{\V\PZ}\xspace}
\newcommand{\ZW}{\ensuremath{\PZ\PW}\xspace}
\newcommand{\ZZ}{\ensuremath{\PZ\PZ}\xspace}
\newcommand{\WW}{\ensuremath{\PW\PW}\xspace}
\newcommand{\WZ}{\ensuremath{\PW\PZ}\xspace}
\newcommand{\Zjets}{\ensuremath{\PZ + \text{jets}}\xspace}
\newcommand{\mVZ}{\ensuremath{m_{\VZ}}\xspace}
\newcommand{\mX}{\ensuremath{m_{\X}}\xspace}
\newcommand{\tZq}{\ensuremath{\PQt\PZ\PQq}\xspace}
\newcommand{\ttV}{\ensuremath{\ttbar\V}\xspace}
\newcommand{\mj}{\ensuremath{m_\mathrm{j}}\xspace}
\newcommand{\mjj}{\ensuremath{m_\mathrm{jj}}\xspace}
\newcommand{\mll}{\ensuremath{m_{\ell\ell}}\xspace}
\newcommand{\EM}{\ensuremath{\Pe\Pgm}\xspace}
\newcommand{\ee}{\ensuremath{\Pe\Pe}\xspace}
\newcommand{\tauo}{\ensuremath{\tau_{1}}\xspace}
\newcommand{\taut}{\ensuremath{\tau_{2}}\xspace}
\newcommand{\tauto}{\ensuremath{\tau_{21}}\xspace}
\newcommand{\B}{\ensuremath{\mathcal{B}}\xspace}
\newcommand{\Mplbar}{\ensuremath{\overline{M}_{\text{Pl}}}\xspace}
\newcommand{\ktilde}{\ensuremath{\widetilde{\kappa}}\xspace}
\newcommand{\DeltaR}{\ensuremath{\Delta R}\xspace}
\newcommand{\tplusX}{\ensuremath{\cPqt + \text{X}}\xspace}
\newcommand{\lljet}{\ensuremath{2\ell + \text{jet}}\xspace}

\cmsNoteHeader{B2G-17-013}

\title{Search for a heavy resonance decaying into a \PZ boson and a \PZ or {\PW} boson in $2\ell2\PQq$ final states at $\sqrt{s}=13\TeV$}

\date{\today}

\abstract{
  A search has been performed for heavy resonances decaying to $\PZ\PZ$ or $\PZ\PW$ in $2\ell2\PQq$ final states, with two charged leptons ($\ell=\Pe,\Pgm$) produced by the decay of a \PZ boson, and two quarks produced by the decay of a {\PW} or \PZ boson.
  The analysis is sensitive to resonances with masses in the range from 400 to 4500\GeV. Two categories are defined based on the merged or resolved reconstruction of the hadronically decaying vector boson, optimized for high- and low-mass resonances, respectively.
  The search is based on data collected during 2016 by the CMS experiment at the LHC in proton-proton collisions with a center-of-mass energy of $\sqrt{s}=13\TeV$, corresponding to an integrated luminosity of 35.9\fbinv.
No excess is observed in the data above the standard model background expectation.
  Upper limits on the production cross section of heavy, narrow spin-1 and spin-2 resonances are derived as a function of the resonance mass, and exclusion limits on the production of \PWpr bosons and bulk graviton particles are calculated in the framework of the heavy vector triplet model and warped extra dimensions, respectively.
}

\hypersetup{pdfauthor={CMS Collaboration},pdftitle={Search for a heavy resonance decaying into a Z boson and a Z or W boson in 2l2q final states at sqrt(s) = 13 TeV},pdfsubject={CMS},pdfkeywords={CMS, physics, BSM, graviton, diboson, ZZ, WZ, boosted, hvt, Wprime, resonance}}

\maketitle

\section{Introduction}

The validity of the standard model (SM) of particle physics is corroborated by a wide set of precise experimental results with an impressive level of accuracy.
Nonetheless, there are several open points where the SM fails to provide an explanation, either for experimental observations, as in the case of the presence of dark matter in the universe, or for theoretical questions, such as the omission of gravity from the SM, and the hierarchy problem.

Several SM extensions addressing the open questions of the SM predict the presence of new heavy particles with an enhanced branching fraction for decays into pairs of vector bosons.
The existence of heavy spin-2 gravitons (\cPG) is predicted in the Randall--Sundrum model with warped extra spatial dimensions (WED)~\cite{Agashe:2007zd,Randall:1999ee,Randall:1999vf}.
In the bulk scenario~\cite{Fitzpatrick:2007qr,Goldberger:1999uk}, the main free parameters are the mass of the first Kaluza--Klein graviton excitation (the bulk graviton mass), and the ratio $\ktilde\equiv \kappa/\Mplbar$, where $\kappa$ is a curvature parameter of the WED metric and $\Mplbar \equiv \Mpl/\sqrt{8\pi}$ is the reduced Planck mass.
The introduction of a spin-1 triplet of \PZpr and \PWpr bosons is described in the heavy vector triplet (HVT) model~\cite{Pappadopulo:2014qza}, which generalizes a large number of explicit models in terms of a small set of parameters: $c_{\mathrm{H}}$, controlling the interactions of the triplet with the SM vector and Higgs bosons; $c_{\mathrm{F}}$, which describes the direct interaction with fermions; and $g_{\mathrm{V}}$, which represents the overall strength of the new vector boson triplet interactions.

A variety of searches for heavy resonances decaying to two vector bosons have been carried out in the past. The most recent results from the CERN LHC~\cite{Khachatryan:2016cfx,Sirunyan:2016cao,Sirunyan:2017nrt,Aaboud:2017ahz,Sirunyan:2018ivv}, with no evidence of signal, have provided stringent upper limits on signal cross sections in these models.

This paper reports on the results of a search for heavy, narrow resonances (collectively indicated as \X) decaying into \llqq final states, with two charged leptons ($\ell=\Pe,\Pgm$) produced by the leptonic decay of a \PZ boson and a pair of quarks produced from the hadronic decay of a vector boson (\V = {\PW} or \PZ). In the narrow-width assumption, the width of the heavy resonance is taken to be small in comparison to the experimental resolution.
Two complementary search strategies are defined to span the mass range $400 < \mX < 4500\GeV$, where \mX is the mass of the heavy resonance.
The first strategy, referred to as the ``high-mass analysis'', is optimized for the range $850 < \mX < 4500\GeV$ by selecting events where the vector bosons have a large Lorentz boost, resulting in the collimation of their decay products.
The high-mass analysis uses dedicated leptonic reconstruction and identification techniques to reconstruct leptons in close proximity to each other in order to retain high signal efficiency, as well as jet substructure techniques to identify the hadronic decay of the {\PW} or \PZ boson into a pair of quarks contained in a single merged reconstructed jet.
For lower resonance masses, the quarks produced by the hadronic decay of the \V boson may be sufficiently separated to be reconstructed as two single narrow jets (dijet).
A second strategy, referred to as the ``low-mass analysis'', is therefore defined in this regime, exploiting dijet reconstruction in addition to the reconstruction of merged jets to retain signal efficiency in the range $400 < \mX < 850\GeV$ for those events in which no merged \V candidate is found.
To increase the signal sensitivity, in the low-mass analysis a categorization based on the flavor of the jets is used, to exploit the relatively large decay branching fraction of the \PZ boson to pairs of b quarks.

This paper is organized as follows: in Section~\ref{sec:mcsimulation}, a description of the data and simulated samples used in the analysis is provided; Section~\ref{sec:detector} briefly describes the CMS detector; Section~\ref{sec:eventreconstruction} provides a description of the event reconstruction; in Section~\ref{sec:evtsel}, the event selection is discussed; Section~\ref{sec:bkg} contains the description of the signal and describes the estimation of the SM background; the systematic uncertainties affecting the analysis are presented in Section~\ref{sec:sys}; and the results of the search for heavy spin-1 and spin-2 resonances are presented in Section~\ref{sec:res}. Finally, results are summarized in Section~\ref{sec:conclusions}.

\section{Data and simulated samples}\label{sec:mcsimulation}

This analysis uses data collected by the CMS detector during proton-proton (\Pp\Pp) collisions at the LHC at $\sqrt{s}=13\TeV$, corresponding to an integrated luminosity of 35.9\fbinv. The events were selected online by criteria that require the presence of at least one electron or muon; these criteria are described in Sec.~\ref{sec:evtsel}.

Simulated signal samples are used in the analysis to optimize the search for the potential production of heavy spin-1 or spin-2 resonances.
For this purpose, signal samples are generated according to the HVT and WED scenarios, respectively. For both scenarios, the samples are generated at leading order (LO) in QCD with the \MGvATNLO~2.2.2 generator~\cite{MADGRAPH}. Two HVT models are considered as benchmarks, ``model A'' and ``model B'', with different values of the three defining parameters described earlier: for ``model A'', $g_{\mathrm{V}}=1$, $c_{\mathrm{H}}=-0.556$, and $c_{\mathrm{F}}=-1.316$, while for ``model B'', $g_{\mathrm{V}}=3$, $c_{\mathrm{H}}=-0.976$, and $c_{\mathrm{F}}=1.024$.

Different resonance mass hypotheses are considered in the range from 400 to 4500\GeV. The resonance width is predicted to be between 0.4 and 2.3\GeV for a \PWpr candidate in HVT model A, and between 14 and 64\GeV for HVT model B, depending on the \PWpr mass hypothesis~\cite{Pappadopulo:2014qza}; in the WED model with $\ktilde = 0.1$, the bulk graviton signal width is predicted to range from 3.6 to 54\GeV~\cite{Oliveira:2014kla}. Since the resonance width is small in comparison with the experimental resolution, for simplicity, the width is taken to be 1\MeV in the simulation.
In the case of the spin-1 \PWpr, the resonance is forced to decay into one \PZ and one {\PW} boson; additionally, the \PZ boson is then forced to decay to a pair of electrons, muons, or tau leptons, while the {\PW} boson is forced to decay into a pair of quarks.
The generated spin-2 bulk graviton is instead forced to decay into two \PZ bosons, one decaying leptonically into any pair of charged leptons, and the other \PZ boson decaying hadronically into a pair of quarks.

Several SM processes yielding final states with charged leptons and jets are sources of background events for the analysis, and corresponding Monte Carlo (MC) simulated samples have been generated and used in the analysis.

The SM production of a \PZ boson in association with quarks or gluons in the final state (\Zjets) represents the dominant background process for the analysis, having topological similarities to the signal because of the presence of a pair of charged leptons and jets.
However, since the quark- and gluon-induced jets are not associated with the decay of a vector boson, the jet mass spectrum is characterized by a smooth distribution and the distribution of the \lljet system invariant mass falls exponentially, in contrast with the peaking distribution expected from the signal in both the jet and \lljet mass spectra.
The \Zjets MC samples are produced with \MGvATNLO at next-to-leading order (NLO), using the FxFx merging scheme~\cite{FXFX} between the jets from matrix element calculations and parton showers, and normalized to the next-to-NLO cross section computed using \FEWZ v3.1~\cite{FEWZ}.

Another important source of SM background arises from processes leading to top quark production.
Simulated samples describing the production of top quark pairs are generated with \MGvATNLO at LO, with the MLM matching scheme~\cite{Alwall:2007fs}.
Single top quark production is also considered; $s$- and $t$-channel single top quark samples are produced in the four-flavor scheme using \MGvATNLO and \POWHEG~v2~\cite{Nason:2004rx,Frixione:2007vw,Alioli:2010xd,POWHEG_ST-st}, respectively, while $\PQt\PW$ production is simulated at NLO with \POWHEG in the five-flavor scheme~\cite{POWHEG_ST-tW}.
Additional top quark background processes, such as the associated production of a \PZ or {\PW} boson with pair-produced top quarks, and the production of $\PQt\PQq\PZ$, are also considered in the analysis and produced at NLO with \MGvATNLO.

The SM diboson production of \VV is an irreducible source of background for the analysis, since the jet mass spectrum will contain a peak from the hadronic decay of {\PW} and \PZ bosons, like the expected jet mass spectrum for the signal; however, this process produces a smoothly falling \lljet invariant mass distribution. The SM production of pairs of vector bosons (\WW, \WZ, and \ZZ) is simulated at NLO with \MGvATNLO.

For all the simulated samples used in the analysis, the simulation of parton showering and hadronization is described by interfacing the event generators with \PYTHIA~8.212~\cite{PYTHIA} with the CUETP8M1~\cite{CUETP8M1} tune, while the parton distribution functions (PDFs) of the colliding protons are given by the NNPDF~3.0~\cite{NNPDF} PDF set.
Additional $\Pp\Pp$ interactions occurring in the same or nearby bunch crossings (pileup) are added to the event simulation, with a frequency distribution adjusted to match that observed in data.
All samples are processed through a simulation of the CMS detector using \GEANTfour~\cite{GEANT4}, and reconstructed using the same algorithms as those for the data collected.

\section{The CMS detector}\label{sec:detector}

The central feature of the CMS apparatus is a superconducting solenoid of 6\unit{m} internal diameter, providing a magnetic field of 3.8\unit{T}. Within the solenoid volume are a silicon pixel and strip tracker, a lead tungstate crystal electromagnetic calorimeter (ECAL), and a brass and scintillator hadron calorimeter (HCAL), each composed of a barrel and two endcap sections. The silicon tracker covers the pseudorapidity range $\abs{\eta} < 2.5$, while the ECAL and HCAL cover the range $\abs{\eta} < 3.0$. Forward calorimeters extend the coverage provided by the barrel and endcap detectors to $\abs{\eta} < 5.2$. Muons are detected in gas-ionization chambers embedded in the steel flux-return yoke outside the solenoid, with detection planes made using three technologies: drift tubes, cathode strip chambers, and resistive-plate chambers.

A more detailed description of the CMS detector, together with a definition of the coordinate system used and the relevant kinematic variables, can be found in Ref.~\cite{Chatrchyan:2008zzk}.

\section{Event reconstruction}\label{sec:eventreconstruction}

The event reconstruction is performed globally using a particle-flow (PF) algorithm~\cite{Sirunyan:2017ulk}, which reconstructs and identifies each individual particle with an optimized combination of information from the various elements of the CMS detector.

The reconstructed vertex with the largest value of summed physics-object $\pt^2$ is taken to be the primary $\Pp\Pp$ interaction vertex. The physics objects chosen are those that have been defined using information from the tracking detector. These objects include jets, the associated missing transverse momentum, which was taken as the negative vector sum of the transverse momentum (\pt) of those jets, and charged leptons.

In the silicon tracker, isolated charged particles with $\pt = 100\GeV$ and $\abs{\eta} < 1.4$ have track resolutions of 2.8\% in \pt and 10 (30)\mum in the transverse (longitudinal) impact parameter \cite{TRK-11-001}.
The energy of charged hadrons is determined from a combination of their momenta measured in the tracker and the matching ECAL and HCAL energy deposits, corrected for zero-suppression effects and for the response function of the calorimeters to hadronic showers. The energy of neutral hadrons is obtained from the corresponding corrected ECAL and HCAL energies.

Electrons are required to be within the range $\abs{\eta} < 2.5$ covered by the silicon tracker, and are reconstructed from a combination of the deposited energy of the ECAL clusters associated with the track reconstructed from the measurements determined by the inner tracker, and the energy sum of all photons spatially compatible with being bremsstrahlung from the electron track.
The identification of electrons is based on selection criteria relying on the direction and momentum of the track in the inner tracker, its compatibility with the primary vertex of the event~\cite{Sirunyan:2017ulk}, and on observables sensitive to the shape of energy deposits along the electron trajectory.
The momentum resolution for electrons with $\pt \approx 45\GeV$ from $\PZ \to \Pe \Pe$ decays ranges from 1.7\% to 4.5\%~\cite{Khachatryan:2015hwa}. It is generally better in the barrel region than in the endcaps, and also depends on the amount of bremsstrahlung emitted by the electron as it traverses the material in front of the ECAL.

Muons are reconstructed in the entire CMS muon system acceptance region of $\abs{\eta}<2.4$ by combining in a global fit the information provided by the measurements in the silicon tracker and the muon spectrometer.
Candidate muons are selected using criteria based on the degree of compatibility of the inner track, which is reconstructed using the silicon tracker only, and the track reconstructed using the combination of the hits in both the tracker and spectrometer.
Further reconstruction requirements include the compatibility of the trajectory with the primary vertex of the event, and the number of hits observed in the tracker and muon systems.
The relative \pt resolution achieved is 1.3--2.0\% for muons with $20 <\pt < 100\GeV$ in the barrel and better than 6\% in the endcaps. The \pt resolution in the barrel is better than 10\% for muons with \pt up to 1\TeV~\cite{Chatrchyan:2012xi}.

Both electrons and muons are required to be isolated from hadronic activity and other leptons in the event.
An isolation variable is defined as the scalar sum of the \pt of charged hadrons originating from the primary vertex, plus the scalar sums of the transverse momenta for neutral hadrons and photons, in a cone of $\DeltaR = \sqrt{\smash[b]{(\Delta\eta)^2 + (\Delta\phi)^2}} <0.3\,(0.4)$ around the electron (muon) direction corrected to account for the contribution from neutral candidates originating from pileup, where $\phi$ is the azimuthal angle in radians.
In the high-mass analysis, a specific muon isolation requirement is implemented to retain signal efficiency for high resonance masses, where the large \PZ boson boost may result in extremely close pairs of muons.
For this reason, muon candidates in the high-mass analysis are retained if they pass an isolation requirement based on the sum of reconstructed \pt of all tracks within $\DeltaR < 0.3$ from the muon trajectory, ignoring tracks associated with other reconstructed muons.

Hadron jets are clustered from particles reconstructed by the PF algorithm using the infrared- and collinear-safe anti-\kt algorithm~\cite{Cacciari:2008gp,Cacciari:2011ma} with distance parameters of 0.4 (AK4 jets) and 0.8 (AK8 jets).
The jet momentum is determined as the vectorial sum of all constituent particle momenta.
Contamination from pileup is suppressed using charged hadron subtraction (CHS) which removes from the list of PF candidates all charged particles originating from vertices other than the primary interaction vertex of the event.
The residual contribution from neutral and charged particles originating from pileup vertices is removed by means of an event-by-event jet-area-based correction to the jet four-momentum.
Identification requirements, based on the estimation of the energy fraction carried by the different types of PF candidates clustered into a jet, along with the multiplicity of the PF candidates, are used to remove jets originating from calorimetric noise.
Corrections to the jet energy are derived from the simulation, and are confirmed with in situ measurements
with the energy balance of dijet, multijet, $\text{photon} + \text{jet}$, and leptonically decaying $\PZ
+ \text{jet}$ events~\cite{Khachatryan:2016kdb}.

A jet grooming technique is used for AK8 jets in this analysis to help identify and discriminate between jets from boosted hadronic \V decays, which we refer to as ``merged jets'', and jets from quarks and gluons.
The AK8 jets are groomed by means of the modified mass drop tagger algorithm~\cite{Dasgupta:2013ihk}, also known as the soft drop algorithm, with angular exponent $\beta = 0$, soft cutoff threshold $z_{\text{cut}} < 0.1$, and characteristic radius $R_0 = 0.8$~\cite{Larkoski:2014wba}.
The soft drop algorithm does not fully reject contributions from the underlying event and pileup.
The mass of the AK8 jet (\mj) is therefore defined as the invariant mass associated to the four-momentum of the soft drop jet, after the application of the pileup mitigation corrections provided by the pileup per particle identification (PUPPI) algorithm~\cite{Bertolini2014}.

Discrimination between AK8 jets originating from vector boson decays and those originating from gluons and quarks is also achieved by the $N$-subjettiness jet substructure variable~\cite{Thaler2011}.
This observable exploits the distribution of the jet constituents found in the proximity of the subjet axes to determine if the jet can be effectively subdivided into a number $N$ of subjets.
The generic $N$-subjettiness variable $\tau_N$ is defined as the \pt-weighted sum of the angular distance of all the $k$ jet constituents from the closest subjet:
\begin{equation}
\tau_N = \frac{1}{d_0} \sum_k p_{\mathrm{T},k} \min( \Delta R_{1,k}, \Delta R_{2,k}, \dots, \Delta R_{N,k} ).
\end{equation}
The normalization factor $d_0$ is defined as $d_0 = \sum_k p_{\mathrm{T},k} R_0$, with $R_0$ the clustering parameter of the original jet.
In this analysis, which aims to select $\V\to\cPq\cPq^{(\prime)}$ decays, the variable that best discriminates V boson jets from those from quarks and gluons is the ratio of the 2-subjettiness to the 1-subjettiness: $\tauto=\taut/\tauo$.
The \tauto observable is calculated for the jet before the grooming procedure, and includes the PUPPI algorithm corrections for pileup mitigation.

For the identification of jets originating from the hadronization of bottom quarks, the combined secondary vertex (CSVv2) algorithm~\cite{btag,CMS-PAS-BTV-15-001} is used, either directly on the AK4 jets or on the AK8 soft drop subjets with CHS pileup mitigation applied.

Only AK4 and AK8 jets reconstructed centrally in the detector acceptance, within $\abs{\eta}<2.4$, are considered in the analysis.

\section{Event selection}\label{sec:evtsel}

Events are selected online by requiring the reconstruction at trigger~\cite{Khachatryan:2016bia} level of at least one charged lepton.
For the high-mass analysis, \pt thresholds of 115 (50)\GeV are used for electrons (muons). No isolation requirements are applied at trigger level, to retain efficiency for high-mass signals, where the large boost expected for the leptonically decaying \PZ boson will cause the two charged leptons to be collimated in the detector.
For the low-mass analysis a larger separation between the leptons is expected because of the lower \pt of the \PZ boson, and isolation requirements are included in the trigger selection, allowing the use of lower lepton \pt thresholds.
The online selection for the low-mass analysis requires at least one electron with $\pt > 25\GeV$ and $\abs{\eta} < 2.1$ passing tight identification and isolation requirements, or at least one muon with $\pt > 24\GeV$ and $\abs{\eta}<2.4$, subject to loose identification and isolation requirements, using the variables described in Ref.~\cite{Khachatryan:2016bia}.

To reconstruct the \PZ boson candidate, at least two well-identified leptons with opposite charge and the same flavor are required to be present in the event.
The leading lepton in the event is required to pass more stringent selection requirements than the online thresholds to avoid inefficiencies induced by the trigger selections.
In the high-mass analysis, the leading (subleading) lepton is required to have $\pt>135\,(35)\GeV$ for electrons, and $\pt>55\,(20)\GeV$ for muons.
Loose isolation and identification requirements are applied to the leptons to retain high signal efficiency. For electrons, we use a set of requirements that have been observed to have an efficiency of about 90\% for both low and high mass points.
For muons, as the CMS standard requirements~\cite{Sirunyan:2018fpa} only have an efficiency of about 65\% for close muons, we instead use a dedicated selection where one of the two muons is allowed to be identified only in the tracker. The isolation variable is calculated removing the contribution of the other muon if it falls within the isolation cone, therefore recovering a signal efficiency of about 90\% for high mass resonances.
For the low-mass analysis, the leading (subleading) lepton is required to have \pt larger than $40\,(30)\GeV$ and to fall in the range $\abs{\eta}<2.1\,(2.4)$.

The selection of the \PZ boson candidate relies on the invariant mass of the dilepton pair, \mll. This is required to satisfy $70 < \mll < 110\GeV$, except for the low-mass analysis in the resolved category (discussed below) where the requirement is $76<\mll<106\GeV$ to enhance the sensitivity to the signal by reducing the nonresonant contribution in the sample with \PQb tagged jets.

Different strategies are used in the low- and high-mass analyses to identify and reconstruct the hadronically decaying \V boson, as described below, to cope with the different \V boson boost regimes expected for low- and high-mass signal candidates.

In the high-mass analysis a merged jet is required in the event, and its mass \mj is used to select the hadronically decaying {\PW} or \PZ.
The signal is expected to be almost fully contained in the mass range $65 < \mj < 105\GeV$, which is thus defined as the signal region (SR).
In order to select candidate signal events, where a heavy massive particle decays into a pair of boosted vector bosons, both the dilepton pair and the leading jet selected in an event are required to have $\pt > 200\GeV$; this is motivated by the \pt spectrum of the \V bosons observed in simulation.
Events are divided into categories depending on the flavor of the charged leptons ({\Pe} or \Pgm) and the value of the jet \tauto variable. As the signal is expected to have lower values of \tauto, two different purity categories are defined: events with $\tauto < 0.35$ are defined as the high-purity (HP) category, while events with $0.35 < \tauto < 0.75$ fall into a low-purity (LP) category, used to retain some sensitivity to signal although a larger amount of background is expected with respect to the HP category.
The $\tauto > 0.75$ region is expected to be dominated by the background, and is therefore not used in the high-mass analysis.
In total, four exclusive categories (from the two purity and two lepton flavor categories) are defined for the high-mass analysis.

In the low-mass analysis, events are divided into two categories depending on whether the two quarks from the hadronic \V decay merge into a single reconstructed jet or can be resolved as two distinct jets. In the merged category, merged jets with $\pt > 200\GeV$ and $\tauto < 0.40$ are selected.
The choice of a looser \tauto selection with respect to the cutoff applied in the HP category of the high-mass analysis is driven by the higher expected signal efficiency for merged events, which are selected in the low-mass analysis using only one \tauto category.
As in the high-mass analysis, the jet mass is required to be in the range $65<\mj<105\GeV$ for the jet to be considered a candidate {\PW} or \PZ boson, which is also defined as the SR for the merged low-mass analysis.
The resolved category contains events that do not contain a merged \V candidate, but instead two AK4 jets, both with $\pt > 30\GeV$ that form a dijet candidate with invariant mass $\mjj > 30\GeV$ and $\pt > 100\GeV$.
In both the merged and resolved cases, the \pt selection is determined by comparing the \pt spectrum of simulated signal events with the expected background.
Both the merged and resolved categories are further split into two \PQb tag categories.
Events in the merged tagged category are required to have at least one subjet satisfying a \PQb tagging requirement corresponding to $\approx$65\% efficiency for b quark identification and $\approx$1\% light-flavor jet mistag rate; events not passing this requirement are placed in the merged untagged category.
For the resolved tagged category, events are required to have at least one jet satisfying the same \PQb tagging requirement used in the merged category; a looser \PQb tag selection is instead required for the other jet, with $\approx$80\% efficiency and $\approx$10\% light-flavor jet mistag rate.
Events failing these requirements fall in the so-called resolved untagged category.
An arbitration procedure is used to select the dijet candidate in case of events containing more than two selected narrow jets: first, if a dijet passing the \PQb tagging requirements is selected in the event, the candidate in the \PQb tag category is chosen; then the dijet candidate closest in mass to the \PZ boson mass is selected as the candidate \V boson.
The signal region for the low-mass resolved category accepts events in the dijet mass range $65<\mjj<110\GeV$.
Eight categories are defined in the low-mass analysis, based on the lepton flavor, the \PQb-tag category, and the merged or resolved reconstruction of the hadronically decaying \V candidate.

\begin{figure*}[!htb]\centering
  \includegraphics[width=\cmsFigWidth]{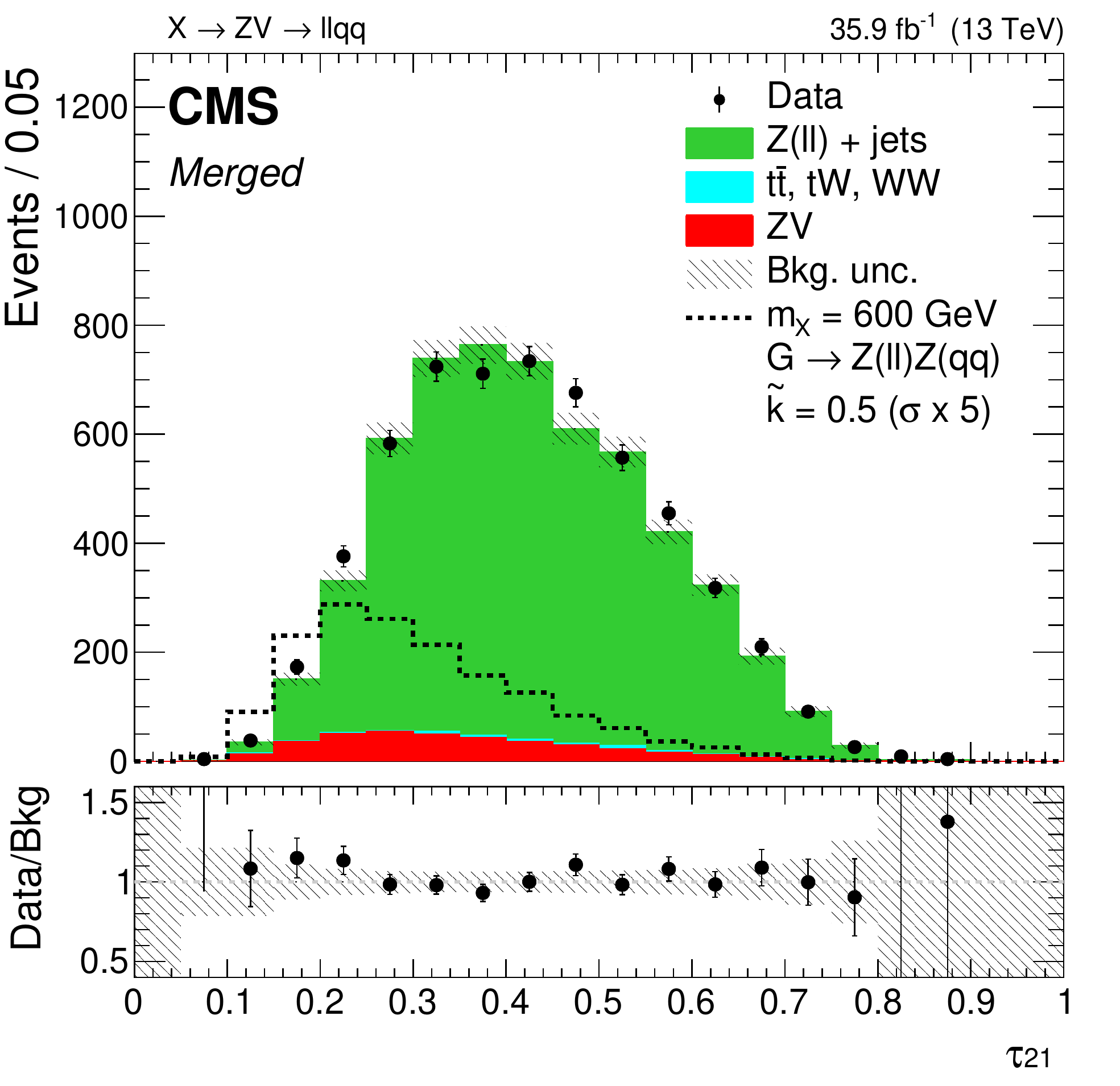}
  \includegraphics[width=\cmsFigWidth]{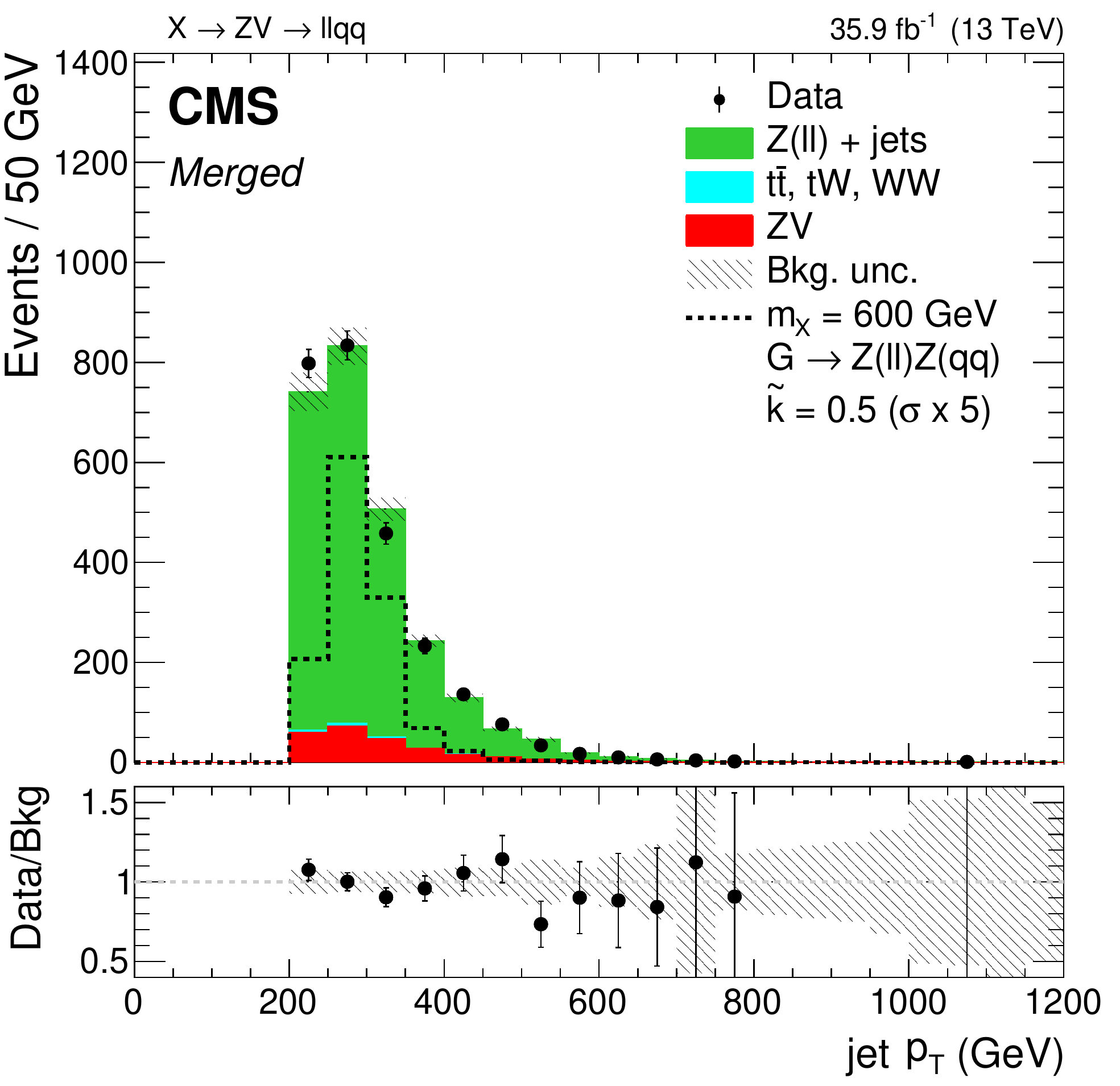}\\
  \includegraphics[width=\cmsFigWidth]{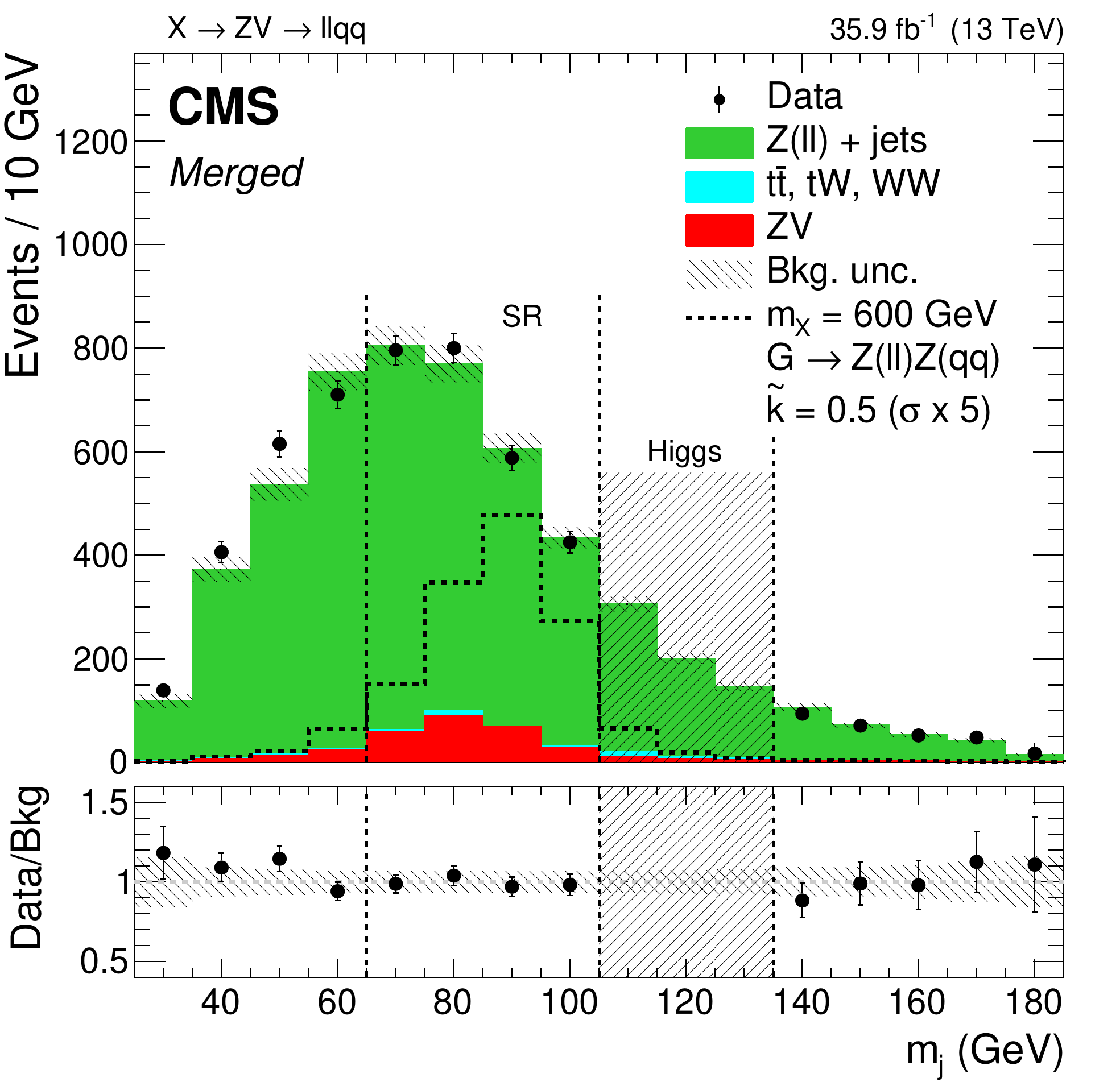}
  \includegraphics[width=\cmsFigWidth]{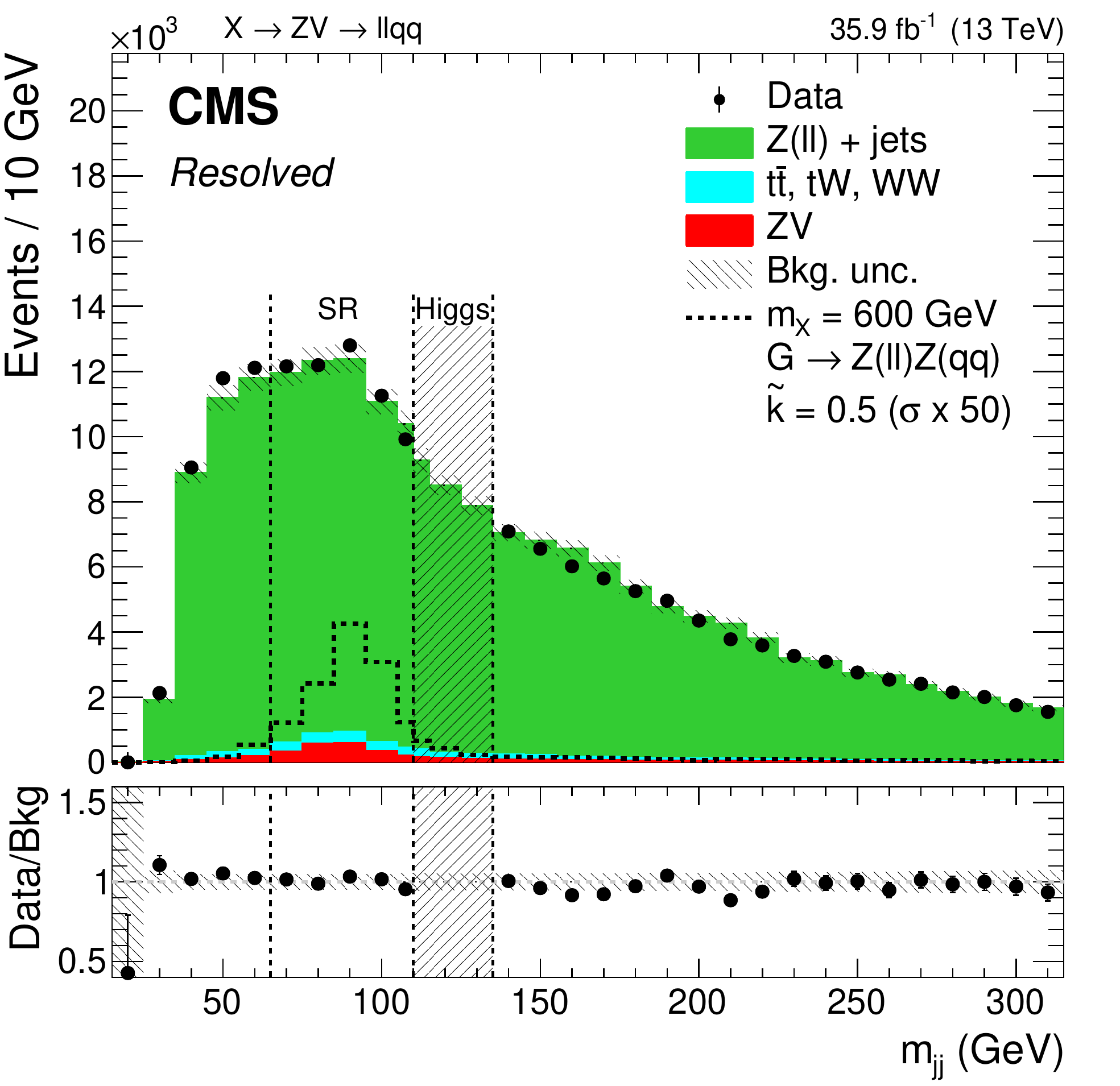}
  \caption{
  Upper row: distribution of the merged \V candidate \tauto (\cmsLeft), where the $\tauto < 0.4$ requirement has been removed, and the jet \pt (\cmsRight) in data and simulation for events in the signal region of the low-mass analysis.
  Lower row: \V candidate \mj (\cmsLeft) and \mjj (\cmsRight) in data and simulation for events in the signal regions of the low-mass search.
  The points show the data while the filled histograms show the background contributions.
  The gray band shows the statistical and systematic uncertainties in the background, while the dashed vertical region (``Higgs'') shows the expected SM Higgs boson mass range, which is excluded from this analysis.
  A 600\GeV bulk graviton signal prediction is represented by the black dashed histogram; for visibility, the signal cross-section is increased by a factor of 5 in the merged category and 50 in the resolved category. With the exception of the jet \pt, which typically peaks at approximately half of the resonance mass, the quantities shown have minimal dependence on the mass of the resonance.
  The background normalization is derived from the final fit to the \mVZ observable in data.\label{fig:vvars_lowmass}
  }
\end{figure*}

The \tauto and merged jet \pt distributions of the \V candidate for events selected in the merged category of the low-mass analysis are shown in Fig.~\ref{fig:vvars_lowmass}, where the \mj and \mjj distributions for events with a \V candidate are also shown for the merged and resolved low-mass analysis categories, respectively.

\section{Background estimation}\label{sec:bkg}

\subsection{High-mass analysis}\label{sec:bkg-high}

The main source of background events in the final state of the analysis arises from the production of a leptonically decaying \PZ boson in association with quark and gluon jets.
A second background source relevant for the analysis is SM diboson production, mainly \ZZ and \ZW, with a leptonically decaying \PZ boson together with a {\PW} or \PZ boson decaying hadronically.
These diboson events are an irreducible background for the analysis, as the mass distribution of the SM \V jet peaks in the same region as the signal.
Finally, top quark production is considered as a source of background in the analysis, despite having a much smaller contribution with respect to other SM backgrounds in the region probed by this analysis, mostly because of the \PZ boson invariant mass selection and the large boost required in the event.

All SM background processes are characterized by a smoothly falling distribution of the invariant mass of the dilepton pair and the jet selected (\mVZ), whereas the signal is instead expected to appear as a narrow peak at a value of \mVZ close to the actual value of the mass of the resonance \mX.

To minimize the dependency on the accuracy of the simulation, the contribution of the dominant background, \Zjets SM production, is estimated using data.
Two signal-depleted regions are defined by selecting events with jet mass outside the \mj signal mass window defined in Section~\ref{sec:evtsel}; these are the sideband (SB) regions.
A lower sideband (LSB) region is defined for events with $30 < \mj < 65\GeV$, close to the SR of the analysis, while a higher sideband (HSB) region contains events with $135 < \mj < 300\GeV$.
The region $105 < \mj < 135\GeV$ is not used in the analysis, to exclude events containing the hadronic decays of a SM Higgs boson, which are targeted in other CMS analyses, such as that described in~\cite{Sirunyan:2018qob}.

The \Zjets background \mVZ shape and normalization are obtained by extrapolation from fits to data in the SB regions.

The \mj distribution for the SM background sources considered in the analysis is modeled by means of analytic functions describing the spectrum of each background in the mass region $30<\mj<300\GeV$.
In the LP category, the \mj spectrum in \Zjets events is described by a smoothly falling exponential distribution, while a broad structure centered around the mass of the {\PW} boson present in the HP category is modeled with an error function convolved with an exponential distribution, which is of particular importance for describing the behavior at large values of \mj.
The peaking structure of the diboson background, originating from the presence of a jet from a genuine {\PW} or \PZ boson in the event, is described in both the LP and HP categories with a Gaussian distribution.
The remaining component of the distribution, consisting of tails extending far from the SR, is modeled in the LP category with an exponential function, similarly to the \Zjets case.
In the HP category, the \VV events are mostly contained in the SR, and the small fraction of events present in the Higgs boson and LSB regions is described with an additional broad Gaussian contribution.
The top quark background (\ttbar, single top quark, \tZq, and \ttV production) is mostly similar in shape to the \Zjets background; in the LP category, in addition to the exponentially falling component, a Gaussian is included to model the top quark peak appearing in the HSB for $\mj\approx170\GeV$.

The expected yield of the \Zjets background in the SR is extracted by a fit of the \mj distribution in the SBs taking into account all background contributions.
The parameters describing the \mj shape and normalization of the subdominant background processes are fixed to those extracted from the simulation.
All the parameters used to describe the \Zjets contribution are left free to float in the fit to the data SBs.
Alternative functions modeling the \mj shape of the main \Zjets background are used to evaluate the impact of the function choice on the signal normalization.

\begin{figure*}[!hbtp]\centering
  \includegraphics[width=\cmsFigWidth]{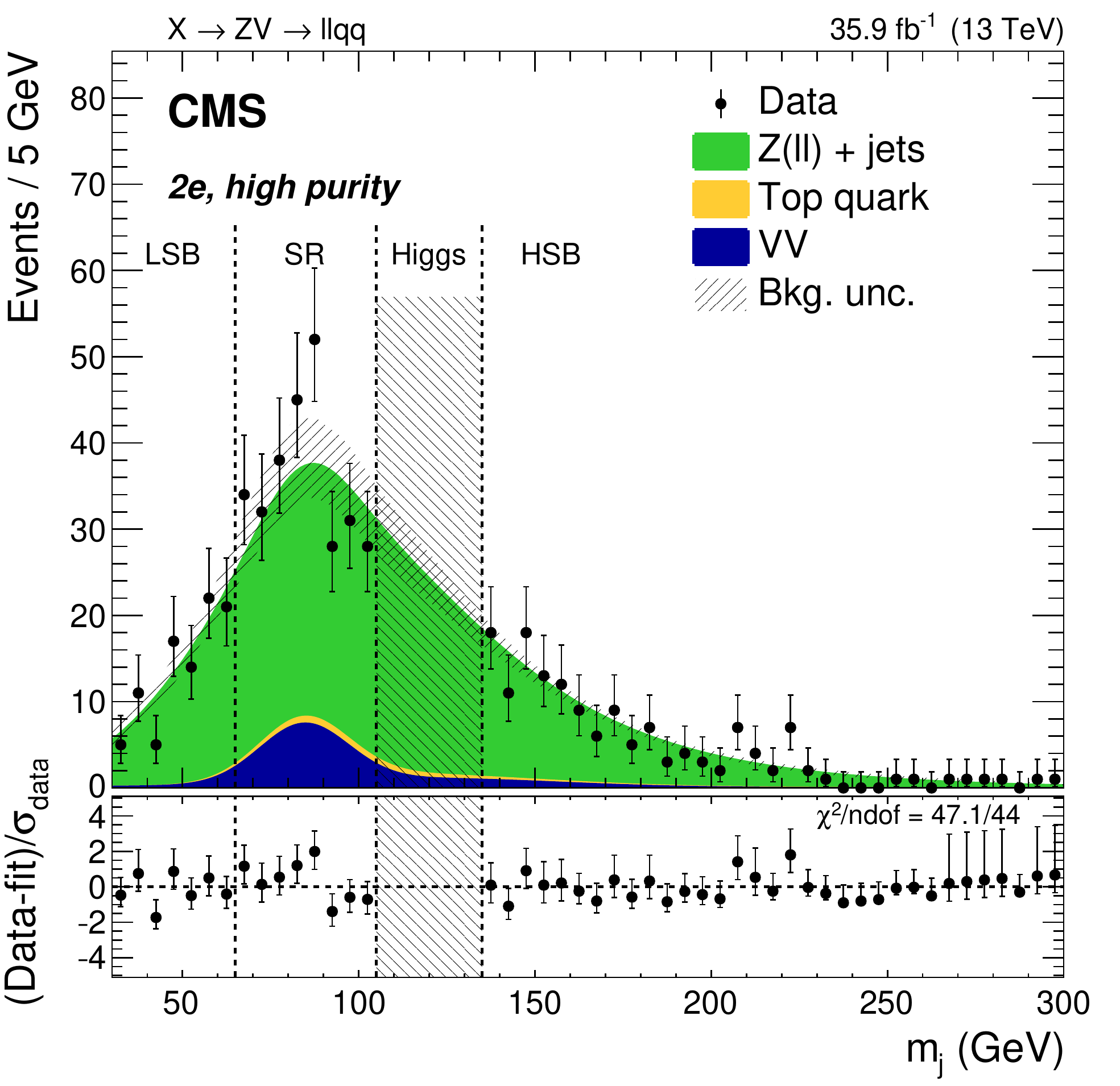}
  \includegraphics[width=\cmsFigWidth]{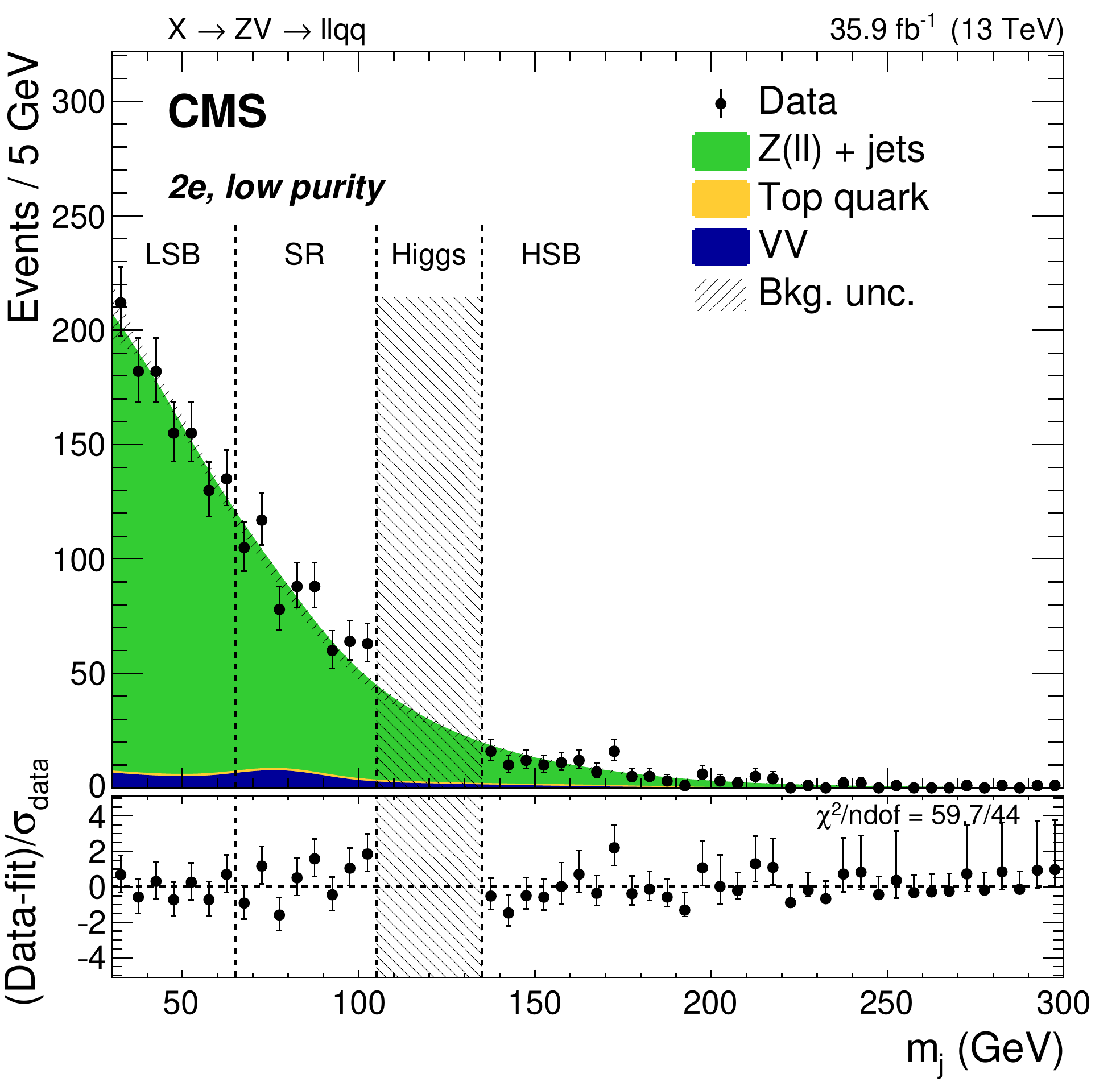}

  \includegraphics[width=\cmsFigWidth]{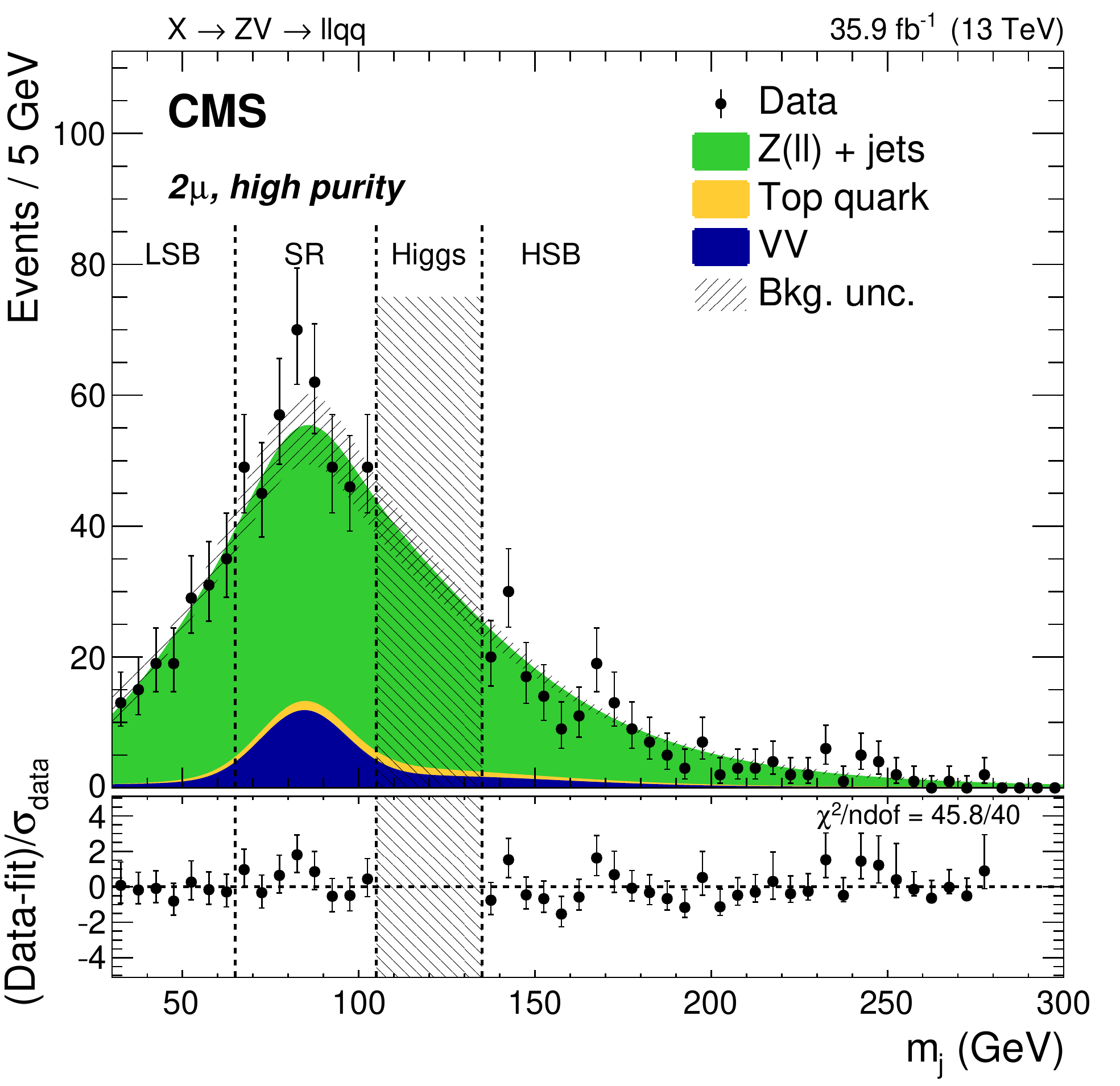}
  \includegraphics[width=\cmsFigWidth]{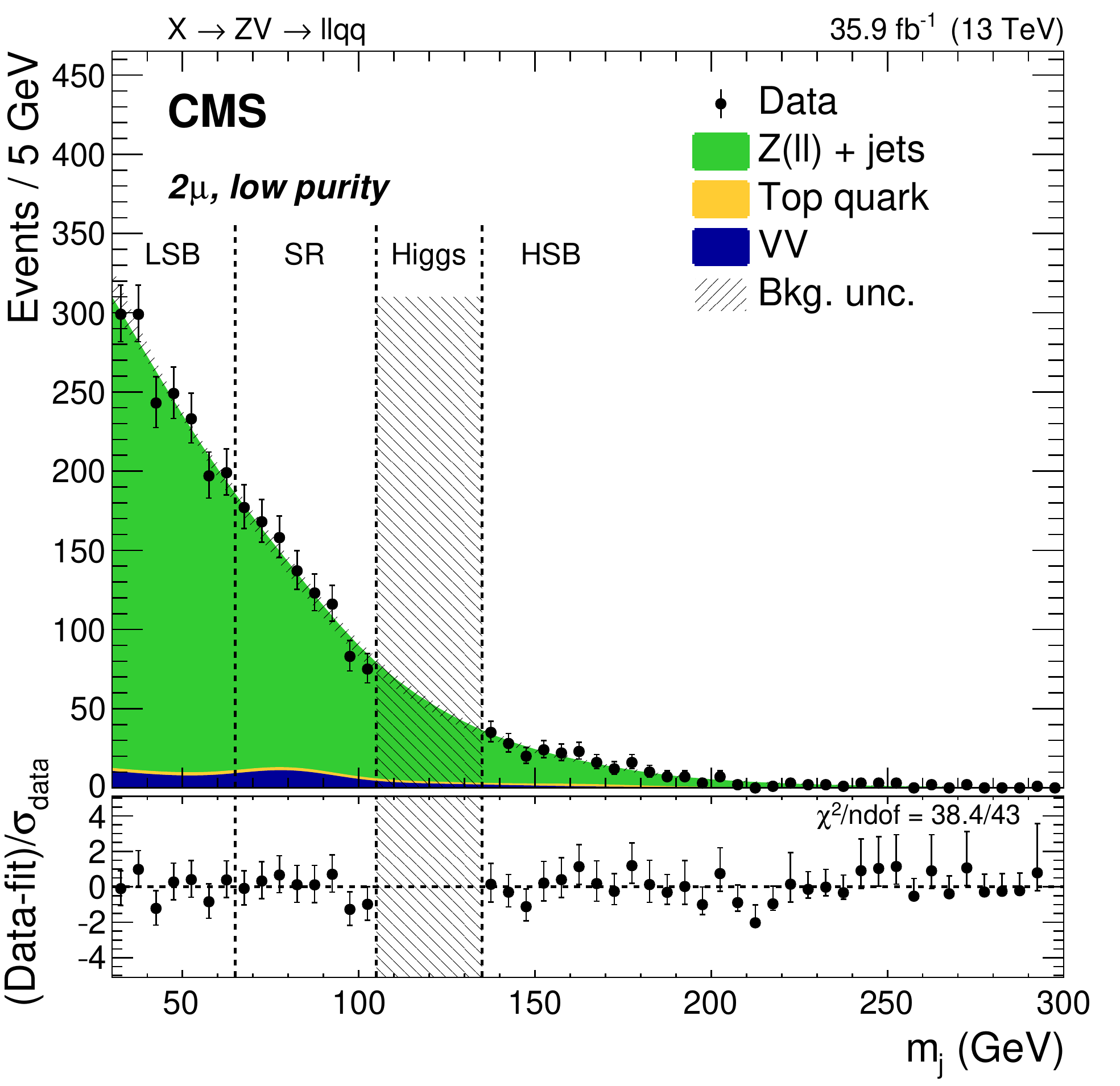}
  \caption{The \mj distributions of the events in data, compared to the expected background shape, for the high-mass analysis in the electron (upper) and muon (lower) channels, and for the high-purity (\cmsLeft) and low-purity (\cmsRight) categories. The expected background shape is extracted from a fit to the data sidebands (\Zjets) or derived from simulation (``top quark'' and ``\VV''). The dashed region around the background sum represents the uncertainty in the \Zjets distribution, while the dashed vertical region (``Higgs'') shows the expected SM Higgs boson mass range, excluded from the analysis. The bottom panels show the pull distribution between data and SM background expectation from the fit, where $\sigma_\text{data}$ is the Poisson uncertainty in the data.}
    \label{fig:mj}
\end{figure*}

The \mj distribution for expected and observed events is shown in Fig.~\ref{fig:mj}.

To describe the shape of the \mVZ variable for the \Zjets background in the SR, the following transfer function is defined from simulation:
\begin{equation}
\alpha(\mVZ) = \frac{f_\text{SR}^{\text{MC},\Zjets}(\mVZ)}{f_\text{SB}^{\text{MC},\Zjets}(\mVZ)},
\end{equation}
where $f_{\text{SR}}^{\text{MC},\Zjets}(\mVZ)$ and $f_{\text{SB}}^{\text{MC},\Zjets}(\mVZ)$ are the probability density functions describing the \mVZ spectrum in the SR and SBs, respectively, of the simulated \Zjets sample.

The shape of the \Zjets background in the SR is then extracted from a simultaneous fit to data in the SBs, and to simulation in both the SR and SBs, to correct the functional form obtained from data using the $\alpha(\mVZ)$ ratio.
The \mVZ shape is described by two-parameter exponential functions for both data and simulation.
The final estimate of the background \mVZ shape predicted in the SR is then given by the following relation:
\begin{equation}
N_{\text{SR}}^{\text{pred}}(\mVZ) = N_{\text{SR}}^{\Zjets} f_{\text{SB}}^{\text{obs},\Zjets}(\mVZ) \alpha(\mVZ) + N_{\text{SR}}^{\text{MC}, \PQt} f_{\text{SR}}^{\text{MC}, \PQt}(\mVZ) + N_{\text{SR}}^{\text{MC}, \VV} f_{\text{SR}}^{\text{MC}, \VV}(\mVZ),
\end{equation}
where $N_{\text{SR}}^{\text{pred}}(\mVZ)$ is the predicted background in the SR and $f_{\text{SB}}^{\text{obs},\Zjets}(\mVZ)$ is the probability distribution function describing the \Zjets background in the SBs.
This is obtained from a fit of the overall background components to data in the SBs, after subtracting the subdominant top quark and \VV components, which are derived from simulation.
The functions $f_{\text{SR}}^{\text{MC},\PQt}(\mVZ)$ and $f_{\text{SR}}^{\text{MC},\VV}(\mVZ)$ are the probability distributions of the top quark and diboson components, respectively, also in this case fixed to the shapes derived from the simulated samples in the SR.
The normalization of the \Zjets background in the SR, $N_{\text{SR}}^{\Zjets}$, is provided by the result of the fit on the \mj data sidebands described above, while the normalization of the top quark and \VV backgrounds, $N_{\text{SR}}^{\text{MC}, \PQt}$ and $N_{\text{SR}}^{\text{MC}, \VV}$, are fixed to the expected yields from simulation.

The $\alpha(\mVZ)$ function accounts for differences and correlations in the transfer process from the SB regions to the SR, and is largely unaffected by uncertainties in the overall \Zjets cross section and distribution shapes.

\begin{figure*}[!hbtp]\centering
  \includegraphics[width=\cmsFigWidth]{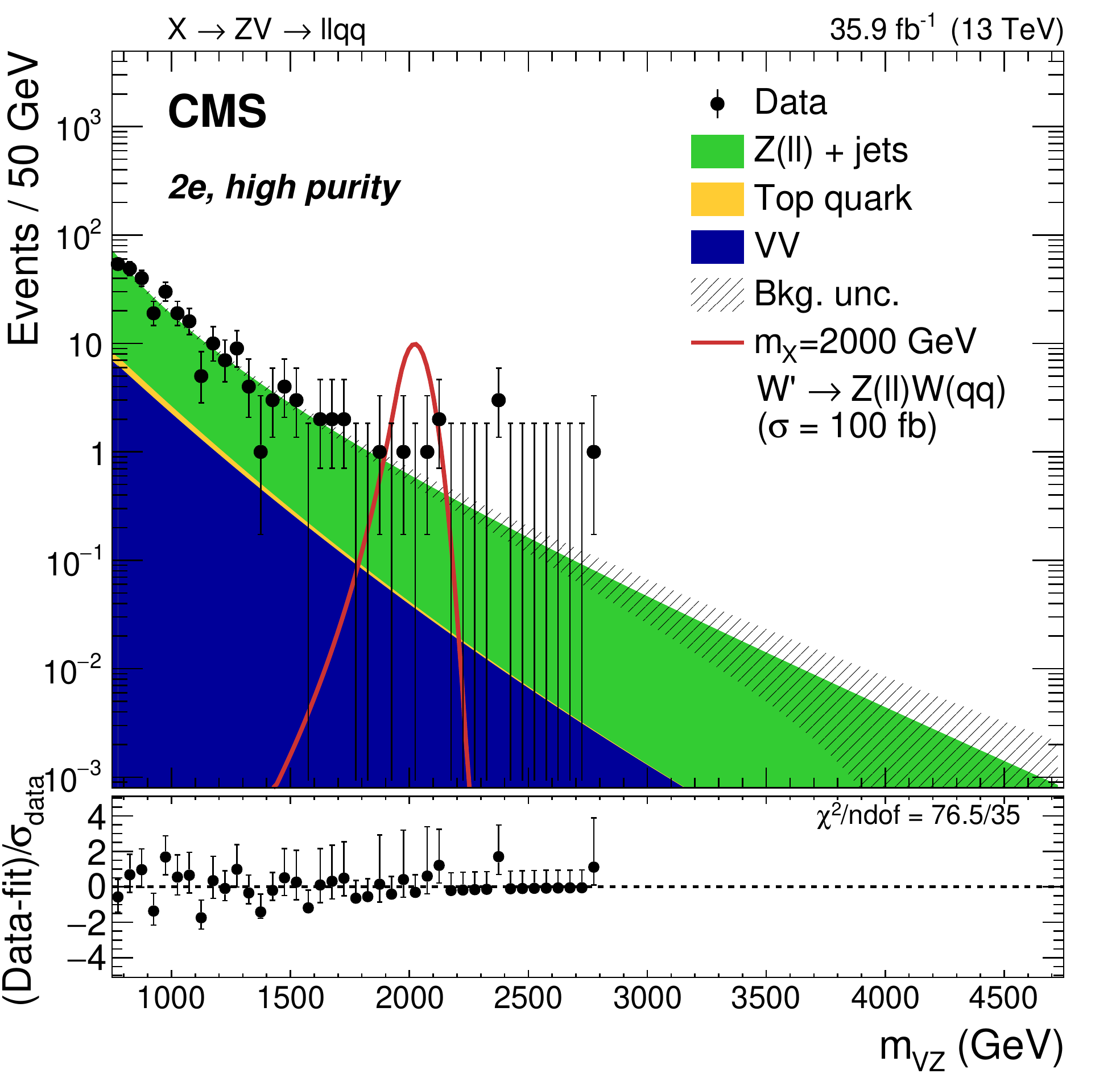}
  \includegraphics[width=\cmsFigWidth]{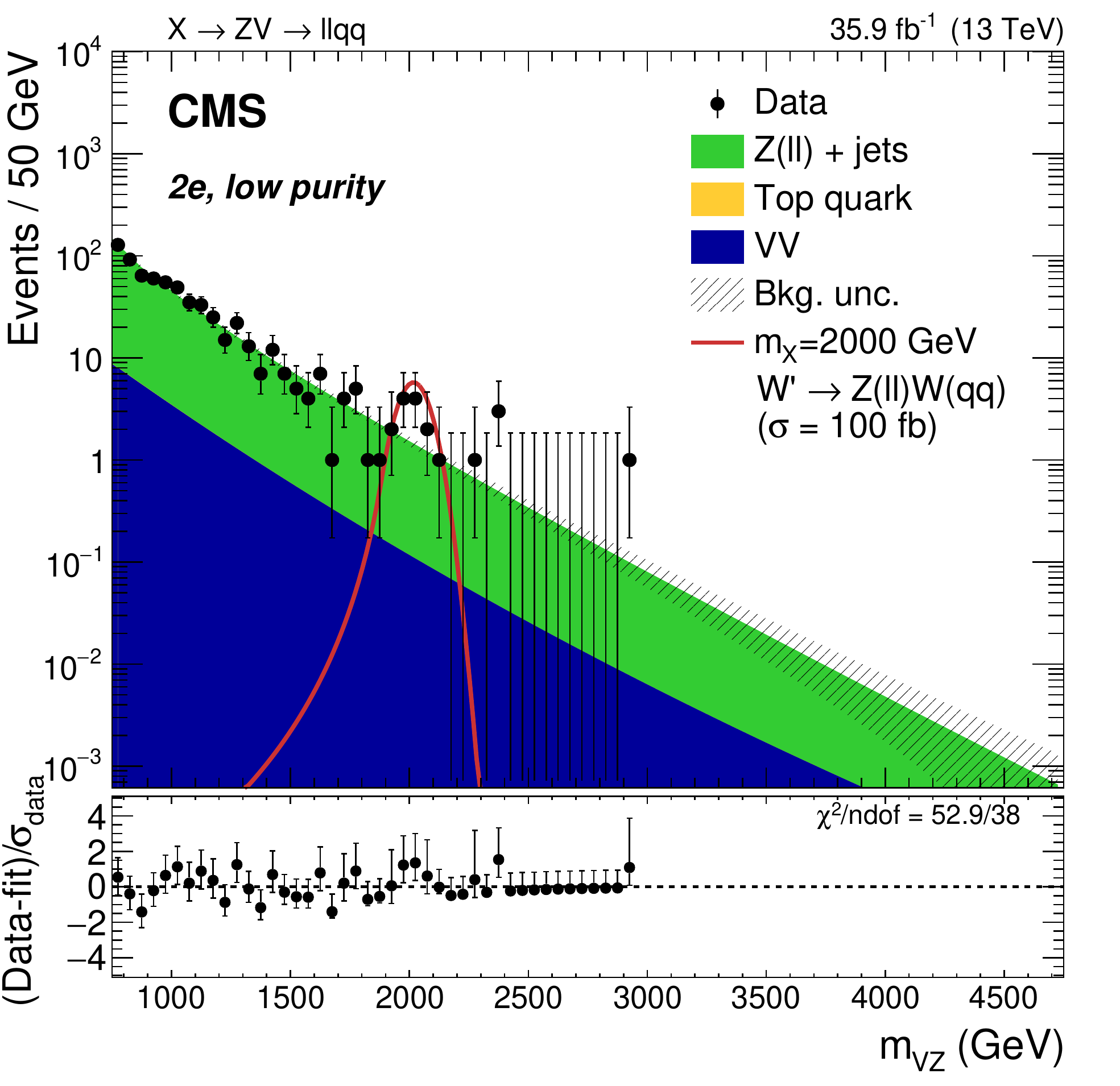}

  \includegraphics[width=\cmsFigWidth]{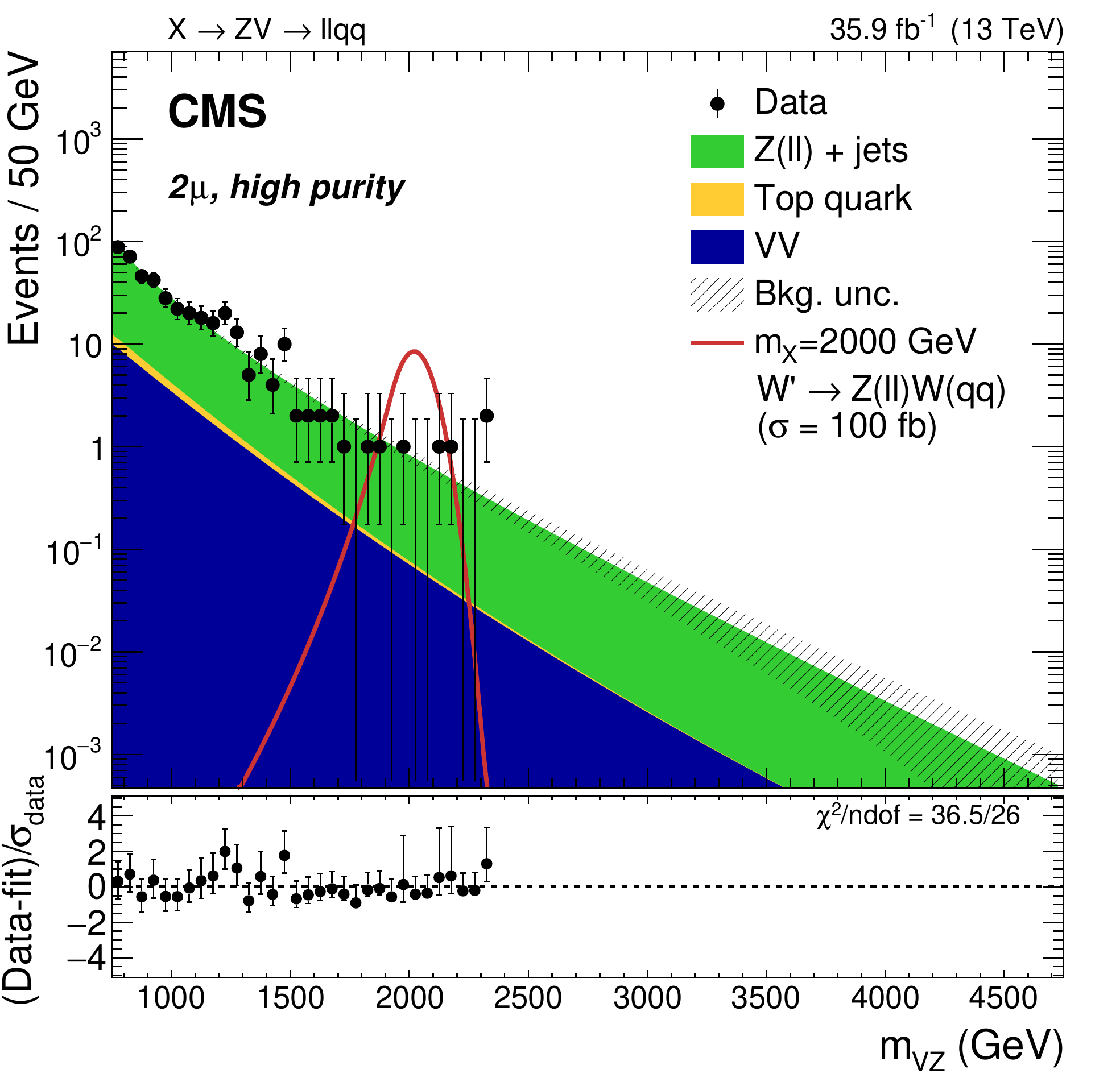}
  \includegraphics[width=\cmsFigWidth]{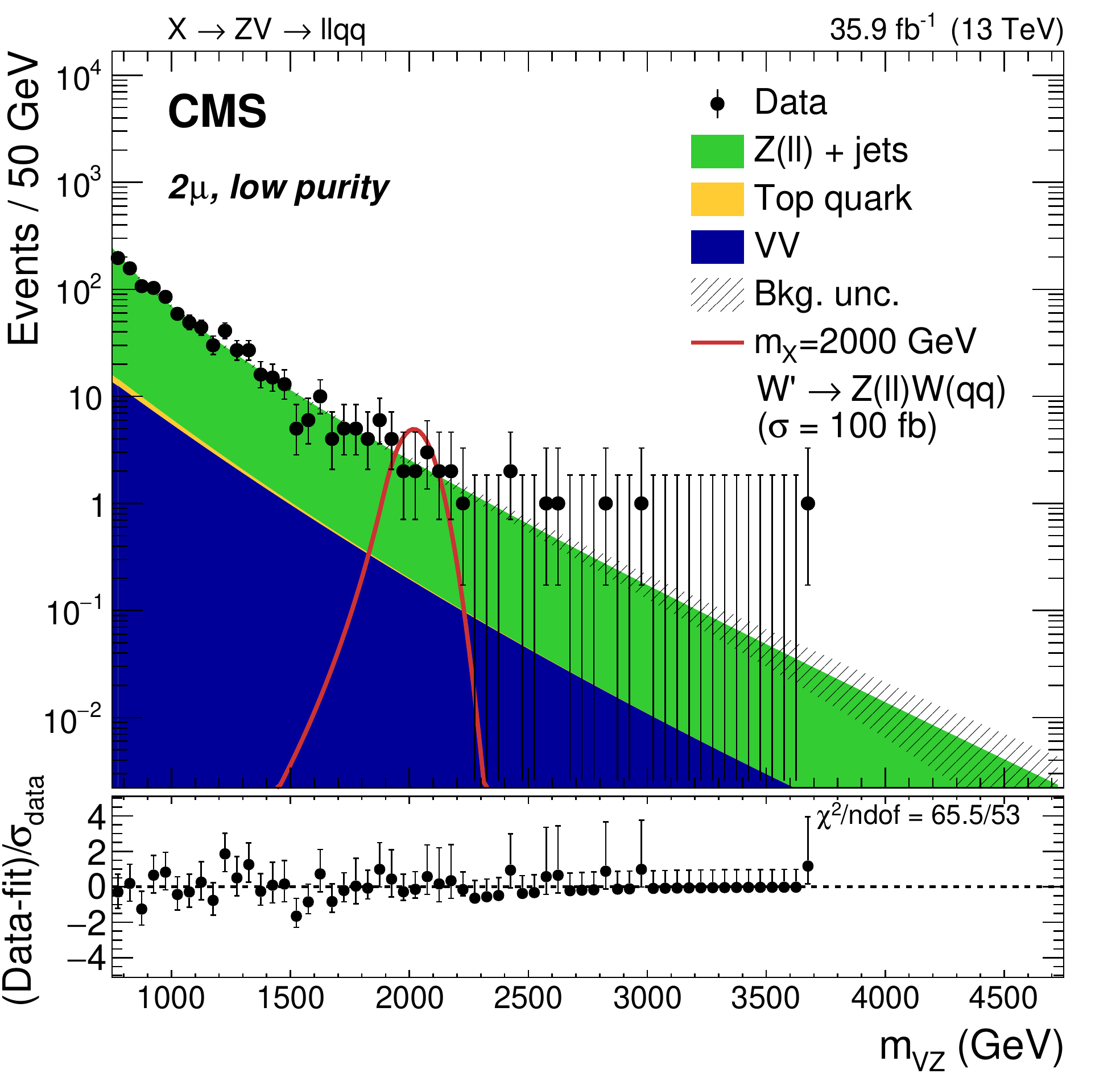}
  \caption{Expected and observed distributions of the resonance candidate mass \mVZ in the high-mass analysis, in the electron (upper) and muon (lower) channels, and separately for the high-purity (\cmsLeft) and low-purity (\cmsRight) categories. The shaded area represents the post-fit uncertainty in the background. The bottom panels show the pull distribution between data and post-fit SM background fit, where $\sigma_\text{data}$ is the Poisson uncertainty in the data.
  The expected contribution from \PWpr signal candidates with mass $\mX=2000\GeV$, normalized to a cross section of 100\unit{fb}, is also shown.\label{fig:mX}}
\end{figure*}

The final \mVZ spectra in the SR are shown in Fig.~\ref{fig:mX}, compared to the expected estimated background.

The validity and robustness of the background estimation method is demonstrated by the agreement observed between the shape and normalization for events selected in an intermediate \mj mass region ($50 < \mj < 65\GeV$), corresponding to the part of the LSB shown in Fig.~\ref{fig:mj} above 50\GeV, and the prediction made using the events in the remaining part of the LSB and the full HSB regions.

The description of the signal \mVZ shape is extracted from  simulated signal samples.
Several signal samples generated with resonance mass ranging from 400 to 4500\GeV in the narrow width approximation are modeled independently for each channel with a Crystal Ball (CB) function~\cite{CBfunction}.
The power-law component of the CB function improves the description of the \mVZ signal distribution by accounting for the small contribution from lower \mVZ tails appearing for high signal masses.
The resolution of the reconstructed \mVZ can be extracted from the Gaussian core width of the CB function, and is estimated to be 2--3.5\% in the electron channel and 3--4\% in the muon channel, depending on the mass of the resonance.

\subsection{Low-mass analysis}\label{sec:bkg-low}

For the low-mass analysis, the \Zjets background is characterized using simulated $\text{Drell--Yan} + \text{jets}$ events.
Because of the limited number of simulated events, the \mVZ distributions in the \PQb-tagged categories are susceptible to sizable statistical fluctuations, which affect the quality of the background modeling.
It has been observed, however, that within simulation uncertainties, the \Zjets mass shape is the same for events with and without \PQb-tagged jets.
Therefore, the \Zjets shape in the \PQb-tagged category is described using the \mVZ shape obtained from the simulation without making any \PQb tag requirements.

Sideband regions are defined depending on the mass of the hadronic \V boson candidate.
The mass ranges $30<\mj<65\GeV$ and $135<\mj<180\GeV$ are used for the merged category, whereas for the resolved event selection the upper mass threshold is raised to 300\GeV to take advantage of the increased number of events in that region.

\begin{figure*}[htb]\centering
\includegraphics[width=\cmsFigWidth]{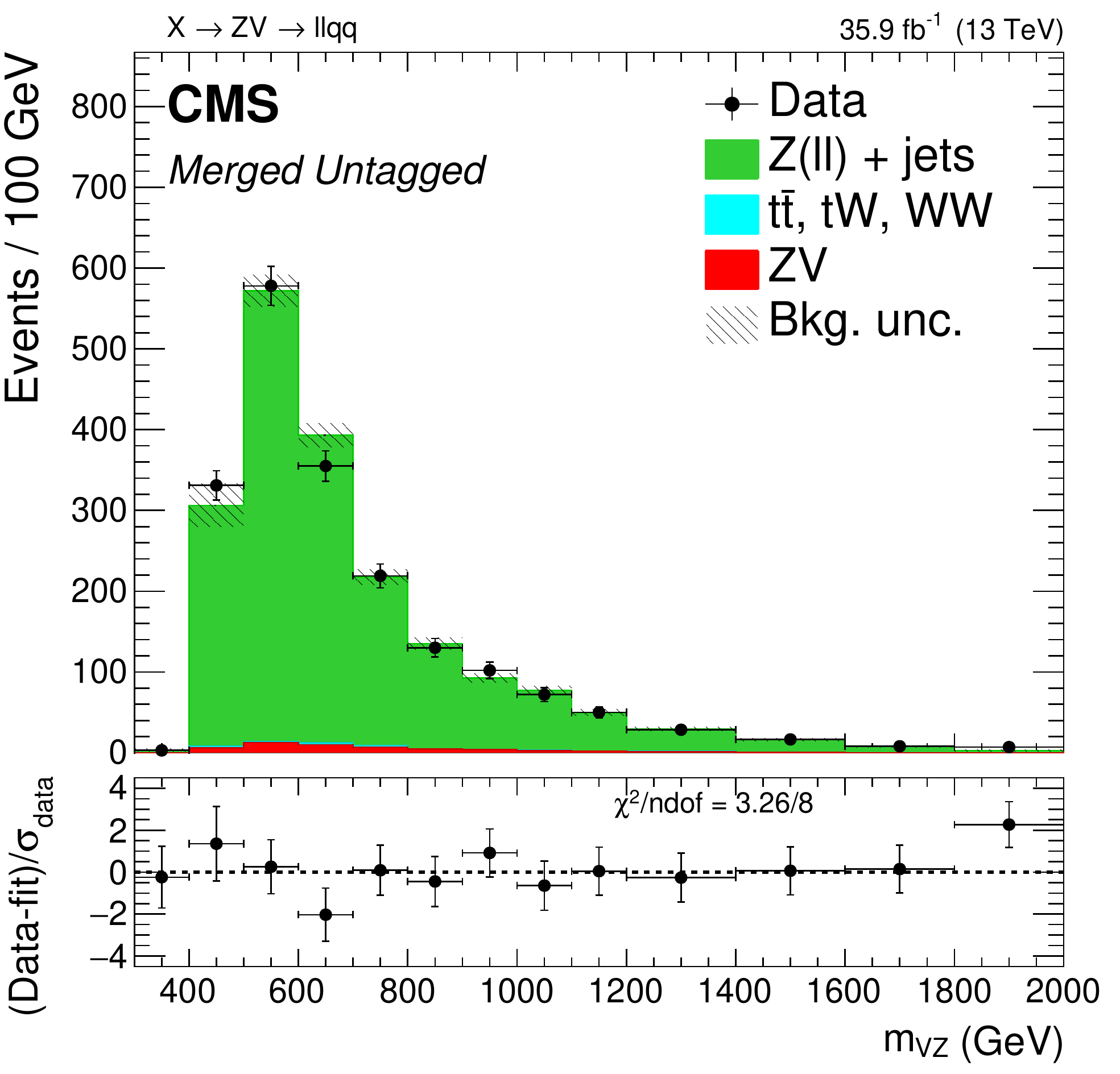}
\includegraphics[width=\cmsFigWidth]{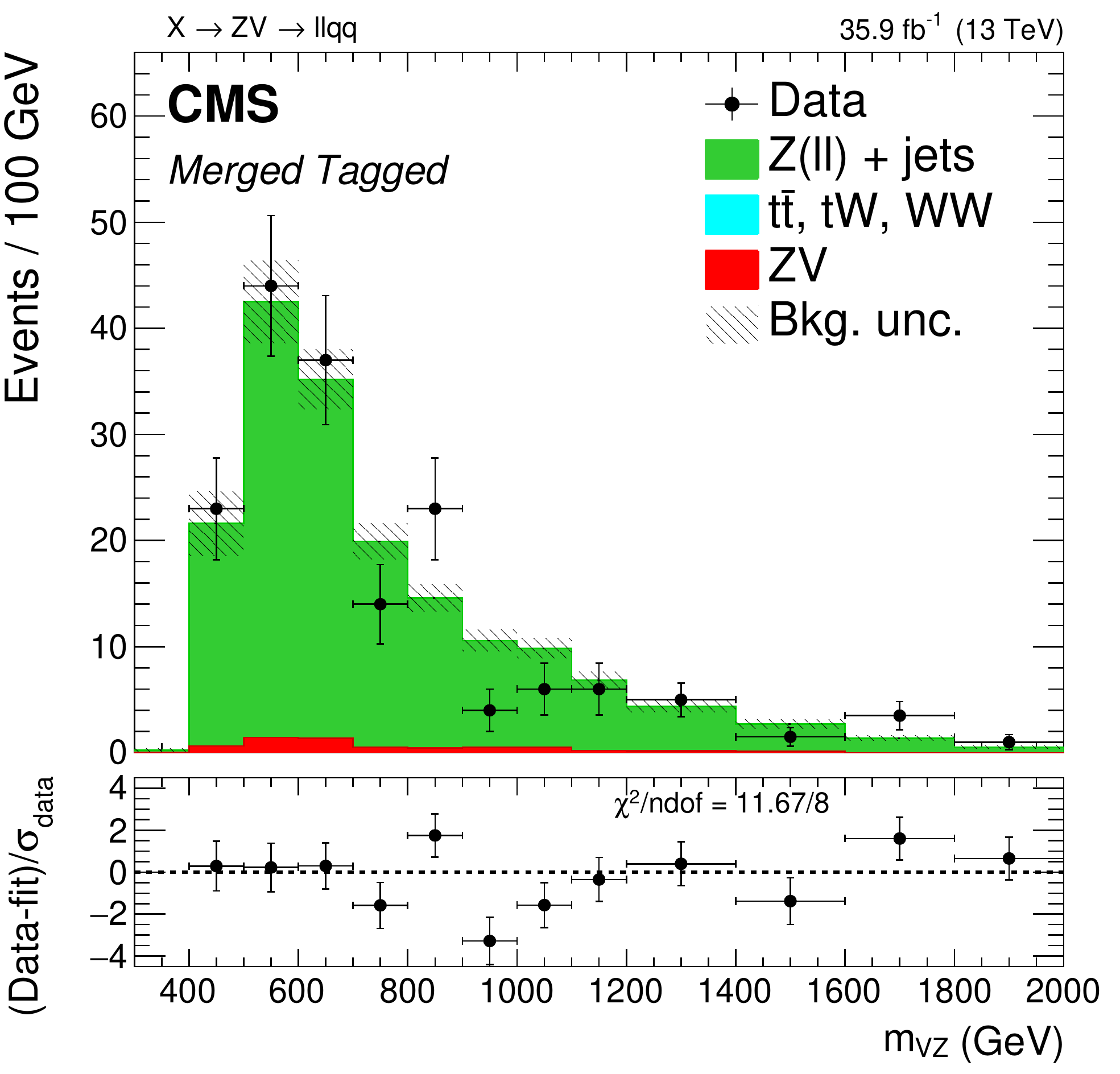} \\
\includegraphics[width=\cmsFigWidth]{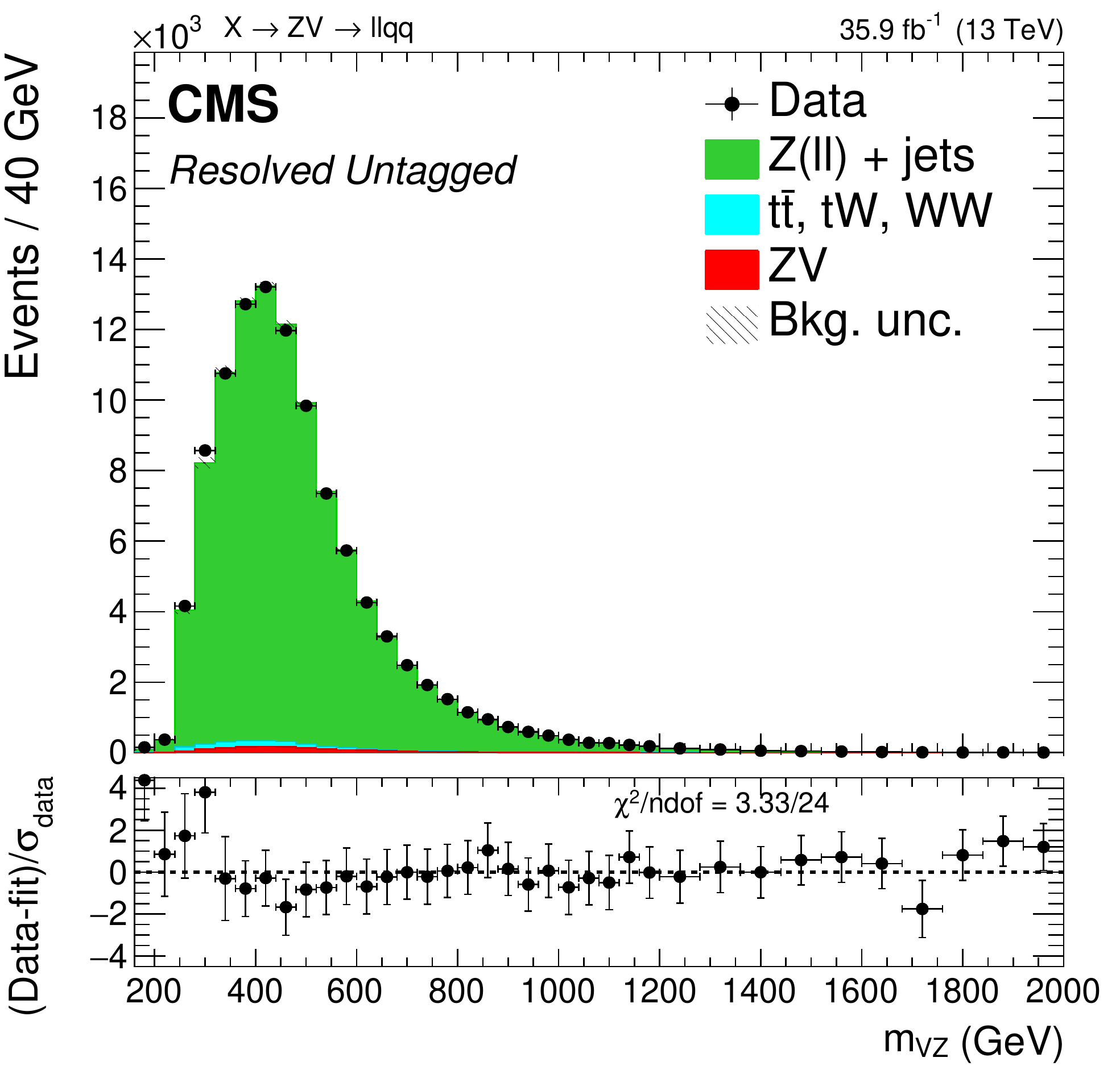}
\includegraphics[width=\cmsFigWidth]{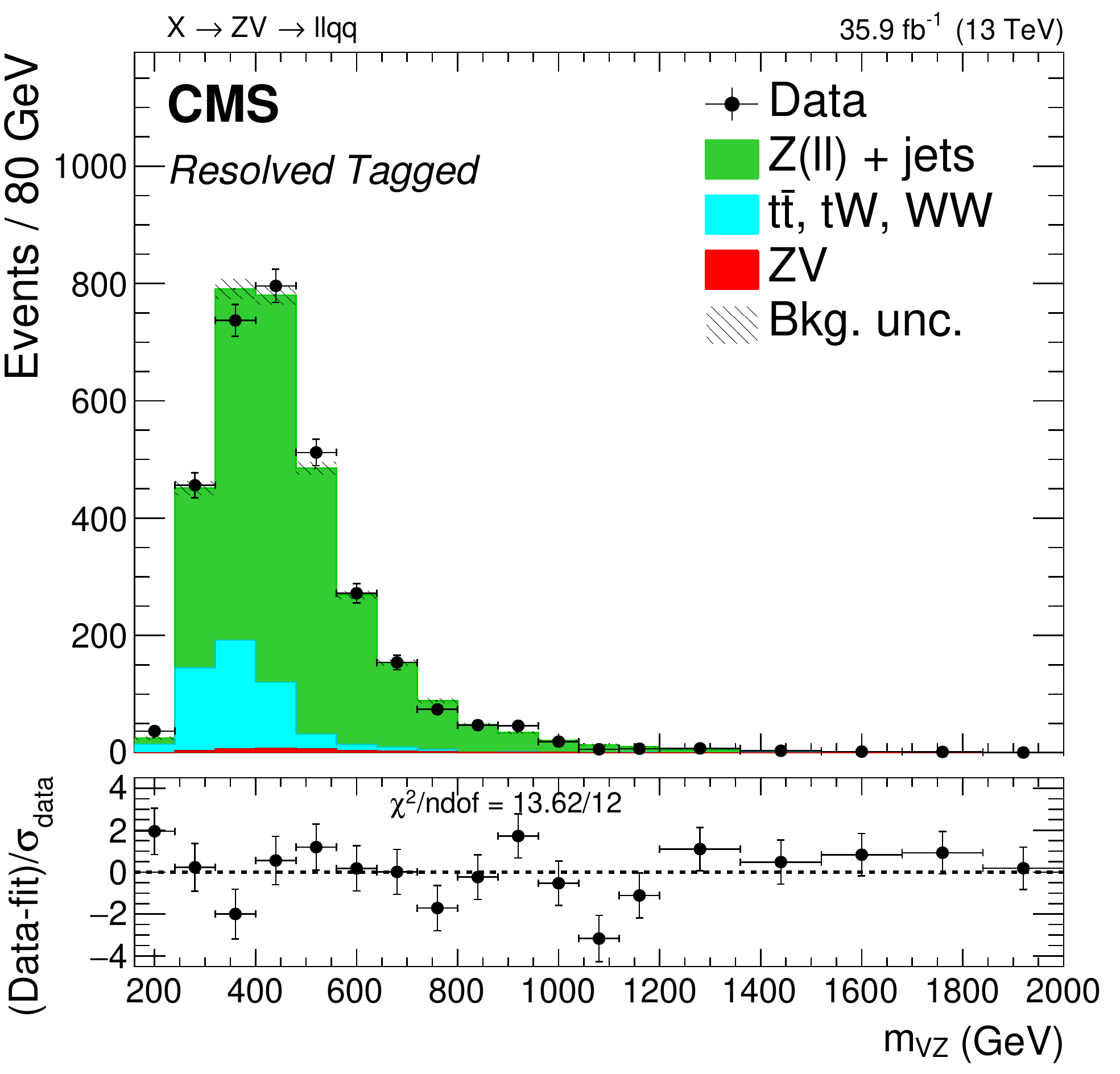}
\caption{
Sideband \mVZ distributions for the low-mass search in the merged \V (upper), resolved \V (lower), untagged (\cmsLeft), and tagged (\cmsRight) categories, after fitting the sideband data alone.
The points show the data while the filled histograms show the background contributions.
Electron and muon categories are combined.
The gray band indicates the statistical and post-fit systematic uncertainties in the normalization and shape of the background.
Larger bin widths are used at higher values of \mVZ; the bin widths are indicated by the horizontal error bars.\label{fig:SB_lowmass}
}
\end{figure*}

In the final fit to the data, the \Zjets background normalization in the SR is constrained by the observed yield in the SBs; this procedure is applied independently to each category.
The shape predictions from the NLO \Zjets simulation are taken as a baseline \mVZ shape in the SR of every category; additionally, a family of linear correction functions:
\begin{equation}
\text{Corr}(\mX, s) = 1 + s (\mX - 500\GeV) / (500\GeV),
\end{equation}
with individual members of the family defined by the slope parameter $s$, is considered.
Figure~\ref{fig:SB_lowmass} shows fits to the SB \mVZ distributions where the slope parameter $s$, allowed to float freely, is constrained by the observed shapes in data.
The two-standard-deviation uncertainties in the fitted linear correction functions, which are in the range from $2\ten{-4}$ to $6\ten{-4}\GeV^{-1}$, depending on the category, are observed to cover the residual shape differences in the SBs.
In the signal region fit of each category, the SB-constrained slope parameter $s$ is treated as a \Zjets shape systematic effect.
In this way the background shape can be corrected to that observed in data.
Statistical uncertainties associated with the simulated \Zjets distributions are also taken into account in the fit. The fits in the merged \V categories include the peaking region of the background; Fig.~\ref{fig:SB_lowmass} shows that the SB data in this particular region are described well by the fit.

Dilepton backgrounds that do not contain a leptonic \PZ boson decay are estimated from data using \EM events passing the analysis selection.
This approach accounts for \ttbar production, $\WW + \text{jets}$, $\Ztott + \text{jets}$, single top quark, and hadrons misidentified as leptons, which we collectively refer to as \tplusX.
The relative yield of \ee and $\Pgm\Pgm$ events with respect to \EM events has been estimated on a top quark--enriched control sample and shown to be consistent with expectations.
Also, the \EM \mVZ distribution was compared with the prediction from simulated background events with symmetric lepton flavor, and found to be in agreement.
The contribution of this \tplusX background is 2\% and 20\% of the total background in the untagged and tagged categories of the resolved analysis, respectively. The merged analysis has a \tplusX contribution of 0.5\% and 1\% in the untagged and tagged categories, respectively.

The diboson background (\ZZ and \ZW, with \Ztoll) is estimated directly from simulation.
The contribution from these events represents 4\% and 5\% of the total background in the untagged and tagged categories of the resolved analysis, respectively, while in the merged analysis it is about 14\% and 16\% in the untagged and tagged categories, respectively.

\begin{figure*}[htb]\centering
\includegraphics[width=\cmsFigWidth]{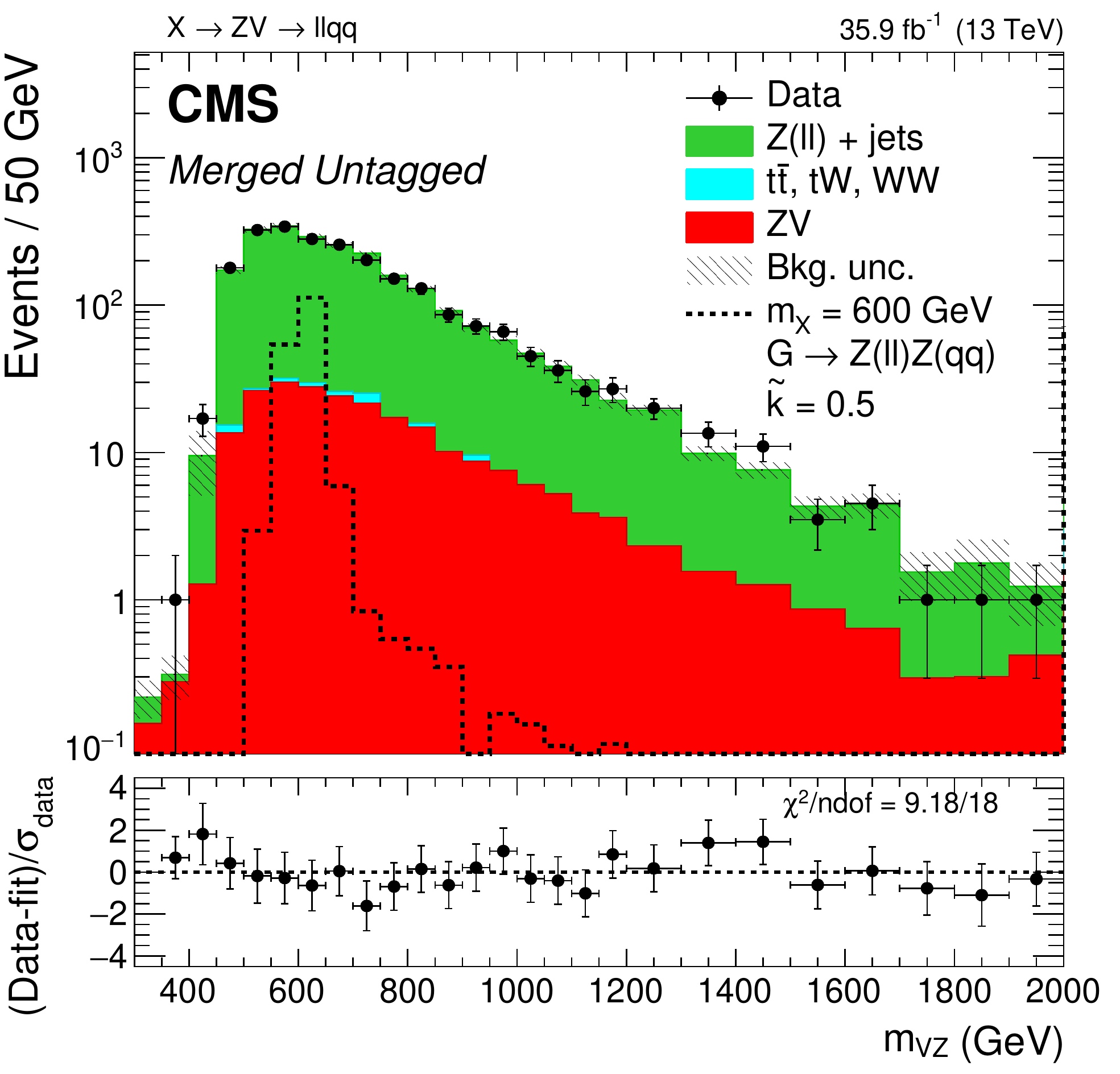}
\includegraphics[width=\cmsFigWidth]{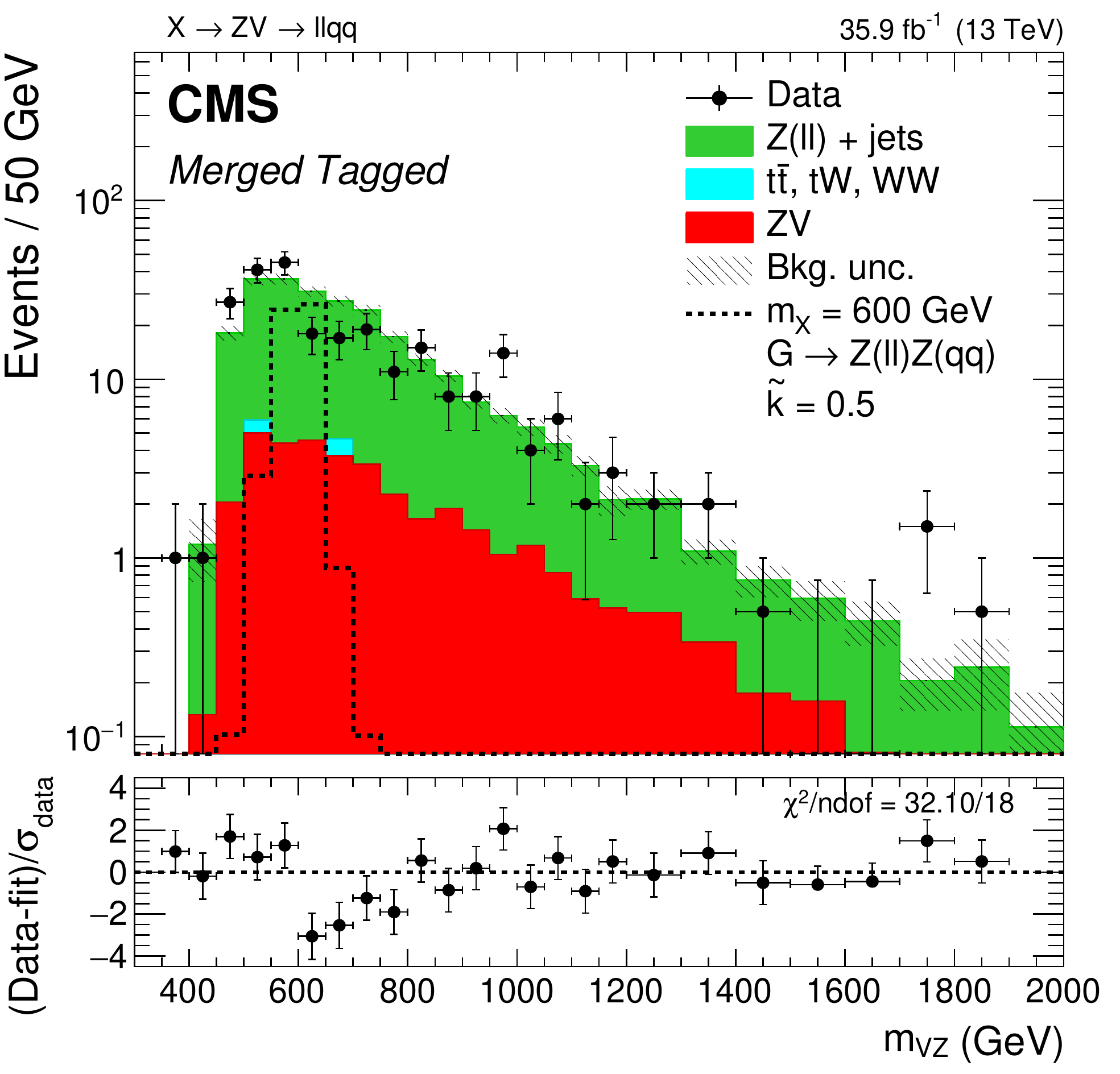} \\
\includegraphics[width=\cmsFigWidth]{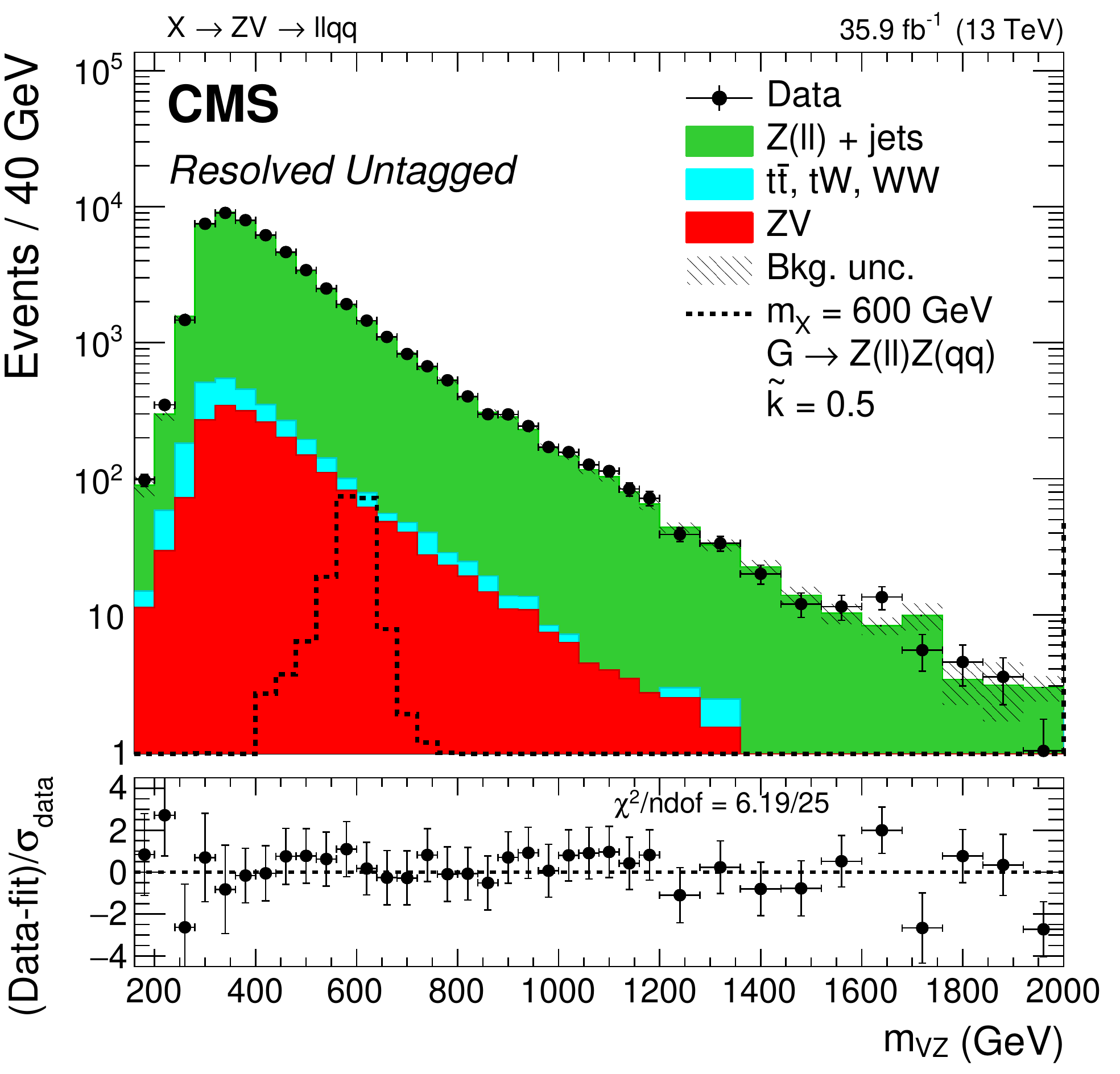}
\includegraphics[width=\cmsFigWidth]{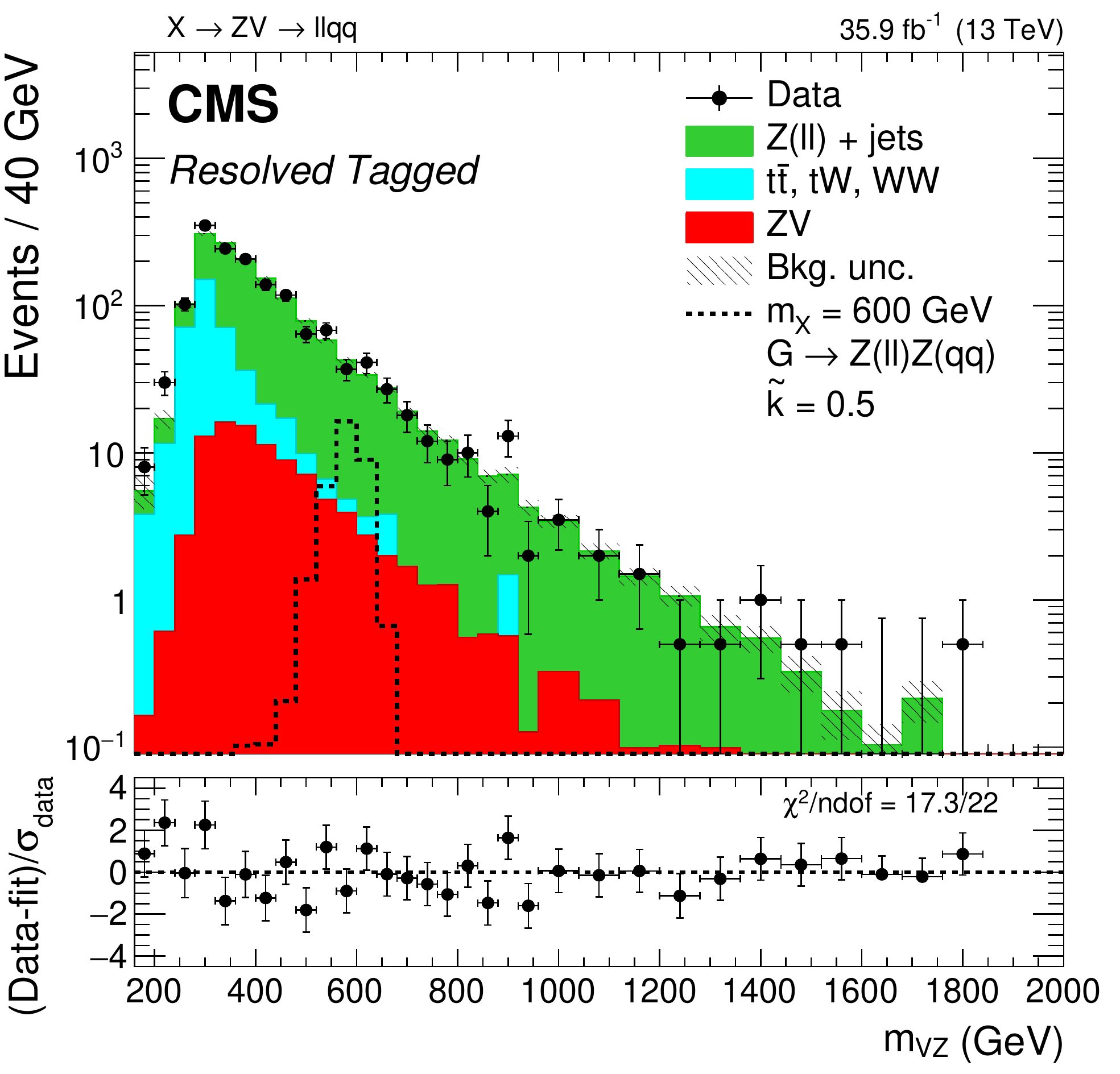} \\
\caption{The signal region \mVZ distributions for the low-mass search, in the merged \V (upper), resolved \V (lower), untagged (\cmsLeft), and tagged (\cmsRight) categories, after fitting the signal and sideband regions.
Electron and muon categories are combined.
A 600\GeV bulk graviton signal prediction is represented by the black dashed histogram.
The gray band indicates the statistical and post-fit systematic uncertainties in the normalization and shape of the background.
Larger bin widths are used at higher values of \mVZ; the bin widths are indicated by the horizontal error bars. \label{fig:SigB_lowmass}
}
\end{figure*}

The \mVZ distributions for the signal region for the merged and resolved categories are depicted in Fig.~\ref{fig:SigB_lowmass}.

\section{Systematic uncertainties}\label{sec:sys}

Several sources of systematic uncertainties influence both the normalization and shape of the backgrounds and signal distributions in the analysis.

In the high-mass analysis, where the \Zjets background component is estimated with data, the main systematic uncertainties in the predicted normalization for the \Zjets background arise from the statistical uncertainties in the fit of the \mj sidebands in data.
Another uncertainty affecting the normalization of the main background is evaluated by taking the difference between the expected \Zjets contribution in the SR obtained by the main function used to describe the \mj spectrum, and an alternative function choice.
An additional normalization uncertainty is related to the choice of the function used to describe the \mj spectrum for the subdominant top quark and \VV backgrounds, evaluated from simulation, and propagated to the \Zjets normalization prediction in the SR.
Overall, the \Zjets normalization uncertainties contribute from 9 to 15\%, depending on the category.
The main shape uncertainties in the \Zjets background are extracted from the covariance matrix of the fit to the \mVZ data SB spectrum, convolved with the uncertainties provided by the $\alpha(\mVZ)$ ratio, via the simultaneous fit procedure described in Section~\ref{sec:bkg-high}.

In the low-mass analysis, to account for background shape systematic effects not explicitly evaluated, data and simulation are compared in the sideband region, and the residual shape difference is treated as an additional uncertainty, resulting in the dominant background shape systematic uncertainty of the low-mass analysis.

The top quark and \VV background components have a systematic uncertainty in the normalization arising from the degree of knowledge of the respective process production cross sections.
The value of the \VV production cross section, taken from a recent measurement by the CMS Collaboration~\cite{Khachatryan:2016txa,Khachatryan:2016tgp}, is assigned an uncertainty of 12\%.
The top quark background uncertainties are estimated differently in the low- and high-mass analyses: in the low-mass analysis, where a dedicated \EM control region is exploited to measure the \tplusX background normalization, a 4\% uncertainty is estimated by comparing the yield of \EM events with $\ee + \Pgm\Pgm$ data; in the high-mass analysis, where the top quark production is taken from simulation, a 5\% uncertainty in the cross section is used, which is extracted from the recent CMS measurement of top quark pair production in dilepton events~\cite{Khachatryan:2016kzg}.

Uncertainties associated with the description in simulation of the trigger efficiencies, as well as the uncertainties in the efficiency for electron and muon reconstruction, identification, and isolation, are extracted from dedicated studies of events with leptonic \PZ decays, and amount to 1.5--3\%, depending on the lepton flavor.
The uncertainties in the lepton momentum and energy scales are taken into account, and propagated to the signal shapes and normalization, with a typical impact on the normalization of about 0.5--2\%, depending on the lepton flavor.

Uncertainties in the jet energy scale and resolution~\cite{CMS-PAS-JME-16-003} affect both the normalization and the shape of the background and signal samples.
The momenta of the reconstructed jets are varied according to the uncertainties in the jet energy scale, and the selection efficiencies and \mVZ signal shapes are reevaluated using these modified samples, resulting in a change of 0.1 to 1.8\%, depending on the jet selection.
The impact of the jet energy resolution is also propagated, and a smaller impact is observed compared with that due to the uncertainty in the energy scale.

The dominant uncertainty in the signal selection efficiency is the uncertainty in the \V boson identification efficiency, corresponding to 11\% (23\%) for the HP (LP) category in the high-mass analysis, and 6\% for the merged category of the low-mass analysis~\cite{Khachatryan2014}.
The V boson identification efficiency, the groomed mass resolution of \V jets, and the related systematic uncertainty are measured in data and simulation in an almost pure selection of semileptonic \ttbar events where boosted {\PW} bosons produced in the top quark decays are separated from the combinatorial \ttbar background by means of a simultaneous fit to the soft drop mass.
The uncertainties in the soft drop mass scale and resolution are propagated to the groomed jet mass, and the impact on the expected selection efficiency of signal and \VV background is taken into account.
An additional uncertainty affecting the signal normalization is included to account for the extrapolation of the uncertainties extracted from a \ttbar sample at typical jet \pt of 200\GeV to higher regimes, estimated from the differences between \PYTHIA~8 and \HERWIG++~\cite{HERWIG} showering models, yielding an uncertainty from 2.5 to 20\% depending on the category.
For the high-mass analysis, the uncertainties in the V boson identification efficiency and the extrapolation are treated as anticorrelated between the low- and high-purity categories.

For the low-mass analysis, one of the largest signal selection uncertainties is the uncertainty in the \PQb tagging efficiency for the tagged categories of the analysis.
The \PQb tagging efficiencies and their corresponding systematic uncertainties are measured in data using samples enriched in b quark content, and their propagation to the signal region of the low-mass analysis produces an uncertainty of up to 4.3\%. The uncertainties in the mistag efficiency are also considered; the uncertainties in the \PQb tagging and mistag efficiencies are treated as anticorrelated between the tagged and untagged categories.

The impact of the uncertainties in the factorization and renormalization scales is propagated both to the normalization and the \mVZ shapes for signal, and for the high-mass analysis to top quark and \VV backgrounds. The corresponding scales are varied by a factor of 2 to measure the effect, resulting in an uncertainty of 2\% for the diboson background normalization and 15\% for top quarks.
The impact on the signal acceptance is evaluated to be 0.1--3\%, depending on the resonance mass and analysis category.

A systematic uncertainty associated with the choice of the set of PDFs used to generate the simulated samples is evaluated by varying the NNPDF~3.0 PDF set within its uncertainties, and its effect is propagated to both the signal and background \mVZ shapes and normalization, resulting in a measured uncertainty of approximately 1\%.

Additional systematic uncertainties affecting the normalization of backgrounds and signal from the contributions of pileup events and the integrated luminosity~\cite{CMS-PAS-LUM-17-001} are also considered and are reported in Table~\ref{table-syst}, together with the complete list of uncertainties considered in the analysis. In the high-mass analysis, the typical total uncertainty in the background normalization is in the range 10--60\%, depending on the signal mass, and it is 1--5\%, depending on the category, in the low-mass analysis.

\begin{table}[!htb]
\centering
\topcaption{
Summary of systematic uncertainties, quoted in percent, affecting the normalization of background and signal samples. Where a systematic uncertainty depends on the resonance mass (for signal) or on the category (for background), the smallest and largest values are reported in the table. In the case of a systematic uncertainty applying only to a specific background source, the source is indicated in parentheses. Systematic uncertainties too small to be considered are written as ``$<$0.1'', while a dash (\NA) represents uncertainties not applicable in the specific analysis category.\label{table-syst}
}
\resizebox{\textwidth}{!}{
\begin{tabular}{l|cc|cc|cc}
                       & \multicolumn{2}{c|}{High-mass} & \multicolumn{2}{c|}{Low-mass} & \multicolumn{2}{c}{Low-mass} \\
                       & \multicolumn{2}{c|}{Merged} 	& \multicolumn{2}{c|}{Merged}   & \multicolumn{2}{c}{Resolved} \\ \hline
Source                 & Background   & Signal       	& Background   & Signal         & Background & Signal     \\
\hline
Electron trigger and ID& \multicolumn{2}{c|}{2.0--3.0}  & \multicolumn{2}{c|}{2.0}      & \multicolumn{2}{c}{2.0} \\
Muon trigger and ID    & \multicolumn{2}{c|}{1.5--3.0}  & \multicolumn{2}{c|}{1.5}      & \multicolumn{2}{c}{1.5} \\
Electron energy scale  & $<$0.1      &  1.0             & 0.8          & 0.1--0.5       & 1.3        & 1.2--2.5   \\
Muon momentum scale    & $<$0.1      &  0.5--2.0        & 0.6          & 0.1--0.4       & 1.4        & 0.2--2.0   \\
Jet energy scale       & 0.1--0.5    &  0.1             & 1.0          & 0.3--0.6       & 1.3        & 0.6--1.8   \\
Jet energy resolution  & $<$0.1      &  $<$0.1          & 0.6          & 0.1            & 0.2        & 0.1--0.2   \\
\PQb tag SF untagged   & \NA         &  \NA             & 0.2          & 0.3--0.4       & 0.1        & 0.6        \\
\PQb tag SF tagged     & \NA         &  \NA             & 2.0          & 2.0--2.3       & 3.8        & 4.1--4.3   \\
Mistag SF untagged     & \NA         &  \NA		        & 0.5          & 0.5--0.6       & 0.4        & 0.2--0.4   \\
Mistag SF tagged       & \NA         &  \NA		        & 1.5          & 0.4--0.6       & 4.3        & 0.5--1.4   \\
SM \VZ production      & 12          &  \NA             & 12           & \NA            & 12         & \NA        \\
SM t quark production  & 5           &  \NA             & 4 (\EM)      & \NA            & 4 (\EM)    & \NA        \\
\V identification (\tauto)        & \NA &  11--23       & 6 (\VZ)      & 6              & \NA        & \NA        \\
\V identification (extrapolation) & \NA &  2.5--20      & \NA          & 2.6--6.0       & \NA        & \NA        \\
\V mass scale          & 0.5--2.5    &  1.0--2.0        & 0.2   (\VZ)  & 0.5--1.1       & \NA        & \NA        \\
\V mass resolution     & 5.5         &  5--6            & 5.6   (\VZ)  & 5.7--6.0       & \NA        & \NA        \\
\Zjets normalization   & 9--15       & \NA              & \NA          & \NA            & \NA        & \NA        \\
Pileup                 & 0.5--4.0    &  0.4        	    & 0.5          &  0.1--0.3      & 0.1        & 0.3--0.5   \\
PDFs                   & 0.3--1.5    &  0.5      	    & \NA          &  1.5--1.6      & \NA        & 0.3--1.1   \\
Renorm./fact. scales   & 2 (\VZ), 15 (\Top) & 1.0--3.0  & \NA          &  0.1--0.3      & \NA        & 0.2--0.3   \\
Integrated luminosity  & \multicolumn{2}{c|}{2.5}       & \multicolumn{2}{c|}{2.5}      & \multicolumn{2}{c}{2.5} \\
\end{tabular}
}
\end{table}

\section{Results and interpretation}\label{sec:res}

Results are extracted separately for the high- and low-mass analyses from a combined maximum likelihood fit of signal and background to the \mVZ distribution, simultaneously in all the categories used in the respective analysis.
An unbinned fit is performed in the high-mass analysis, while a binned fit is performed in the low-mass one; this choice is determined by the fact that in the high-mass analysis, the signal and background shapes are described with analytical functions, while in the low-mass analysis, the background shapes are described by binned histograms.
The systematic uncertainties discussed in Section~\ref{sec:sys} are included as nuisance parameters in the maximum likelihood fit, and the background-only hypothesis is tested against the combined background and signal hypothesis~\cite{Read:2002hq,Junk:1999kv}.

The largest excess of events with respect to the background-only hypothesis, with a local significance of 2.5 standard deviations, is observed in the vicinity of $\mX\approx1.2\TeV$, and arises predominantly from a localized excess of events in the dimuon HP category of the high-mass analysis.

The limit at 95\% confidence level (\CL) on the signal cross section for the production of a heavy spin-1 or spin-2 resonance is set using the asymptotic modified frequentist method ($\text{CL}_\mathrm{s}$)~\cite{Read:2002hq,Junk:1999kv,Cowan:2010js,CMS-NOTE-2011-005}.

The results of the low- and high-mass analyses should agree for the intermediate mass range 800--900\GeV, which is accessible to both strategies with similar expected efficiencies for signal candidates.
The results of the analysis are therefore presented based on the low-mass strategy up to resonance masses $\mX \leq 850\GeV$, and based on the high-mass analysis for $\mX \geq 850\GeV$.
At the intermediate mass point $\mX = 850\GeV$, the results of both strategies are presented, and the expected limits at 95\% \CL of the low- and high-mass analyses on the signal cross sections are found to be in agreement within 3 and 6\% for the \PWpr and bulk graviton signal model, respectively.

\begin{figure}[!htb]\centering
    \includegraphics[width=1.15\cmsFigWidth]{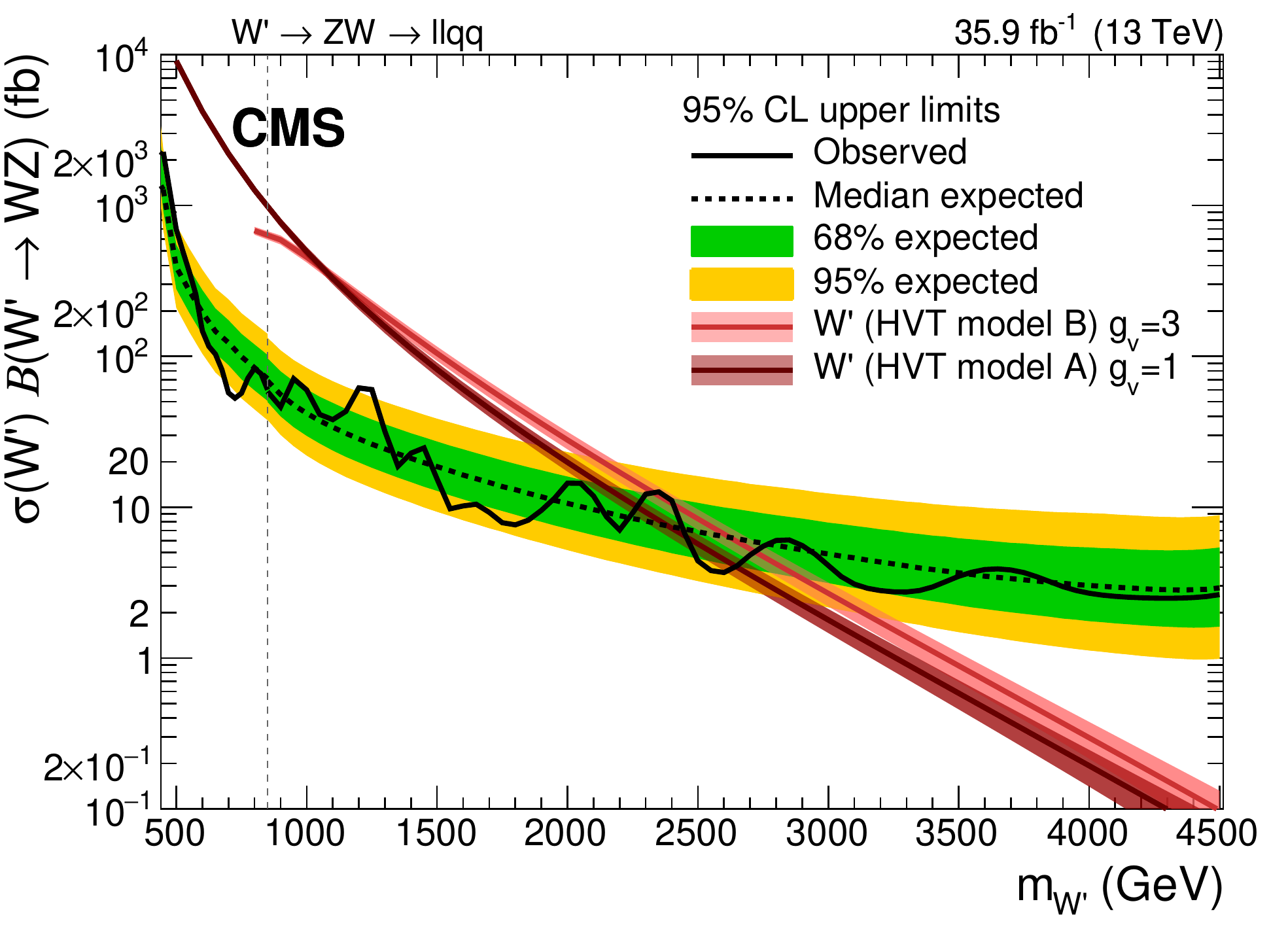}
    \includegraphics[width=1.15\cmsFigWidth]{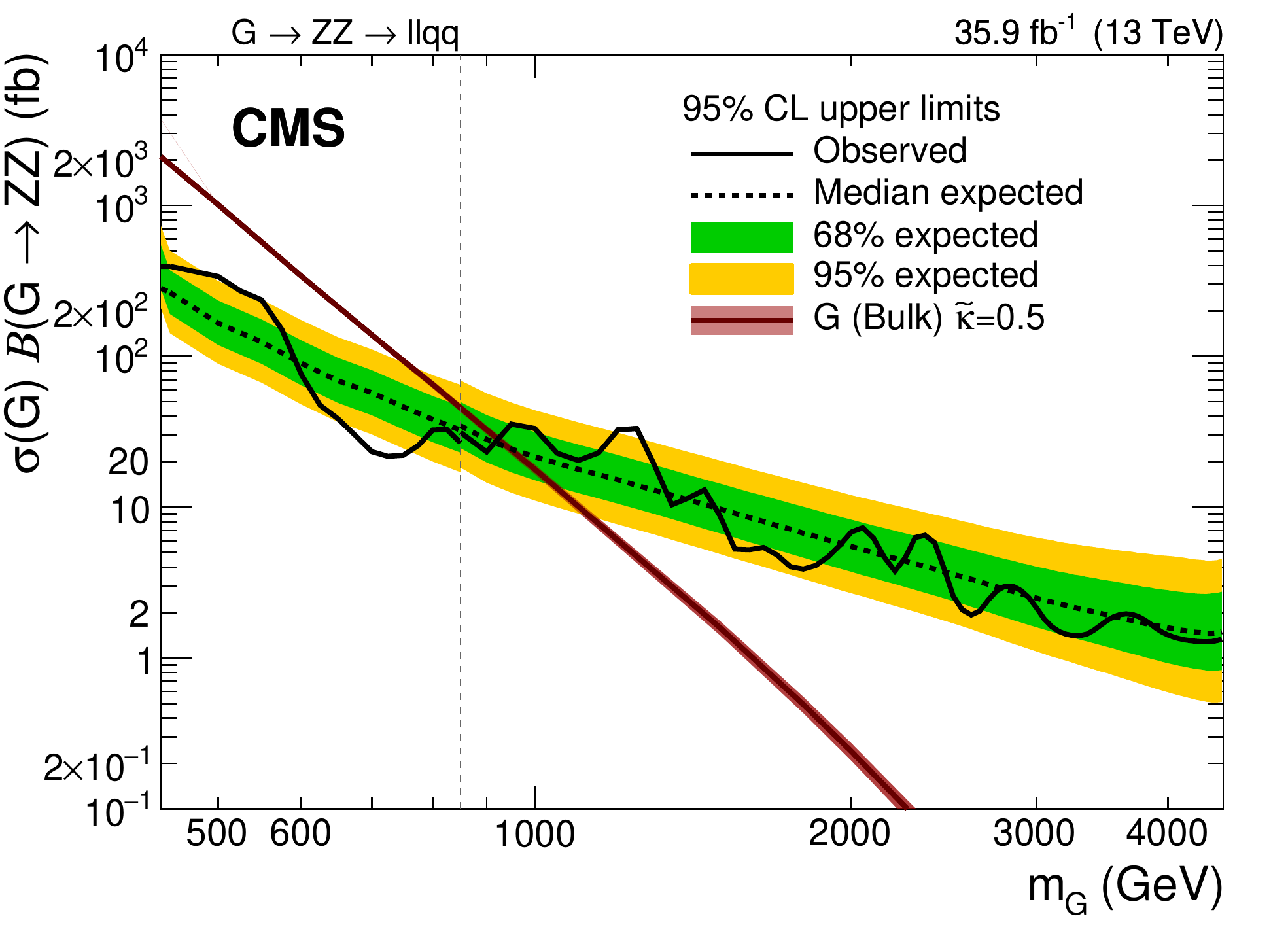}
    \caption{Observed and expected 95\% \CL upper limit on $\sigma_{\PWpr} \B(\PWpr\to\ZW)$ (\cmsLeft) and $\sigma_{\cPG} \B(\cPG\to\ZZ)$ (\cmsRight) as a function of the resonance mass, taking into account all statistical and systematic uncertainties. The electron and muon channels and the various categories used in the analysis are combined together. The green (inner) and yellow (outer) bands represent the 68\% and 95\% coverage of the expected limit in the background-only hypothesis. The dashed vertical line represents the transition from the low-mass to the high-mass analysis strategy. Theoretical predictions for the signal production cross section are also shown: (\cmsLeft) \PWpr produced in the framework of HVT model A with $g_\mathrm{v}=1$ and model B with $g_\mathrm{v}=3$; (\cmsRight) {\cPG} produced in the WED bulk graviton model with $\ktilde=0.5$.}
  \label{fig:limit}
\end{figure}

\begin{figure}[!htb]\centering
    \includegraphics[width=1.15\cmsFigWidth]{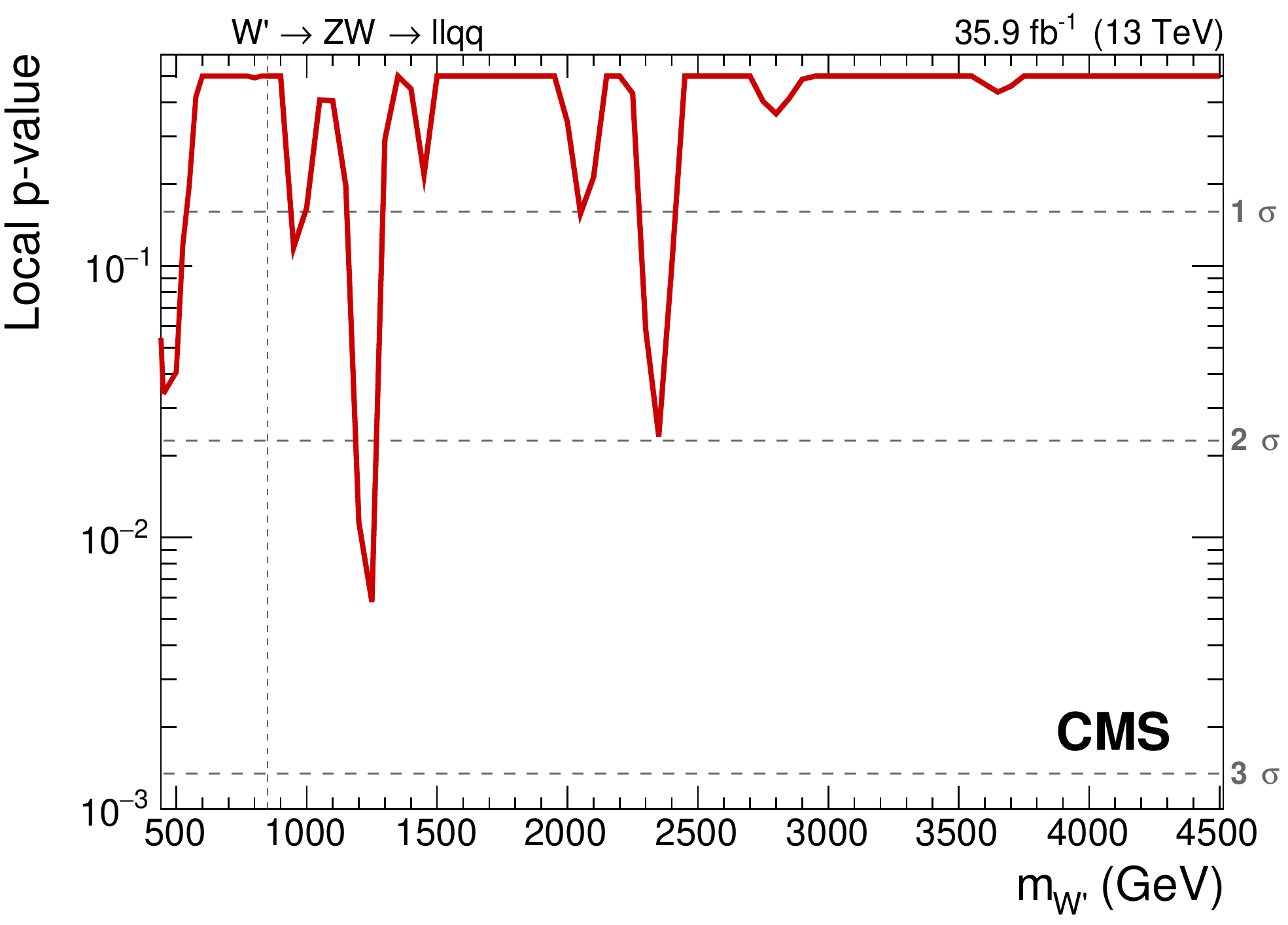}
    \includegraphics[width=1.15\cmsFigWidth]{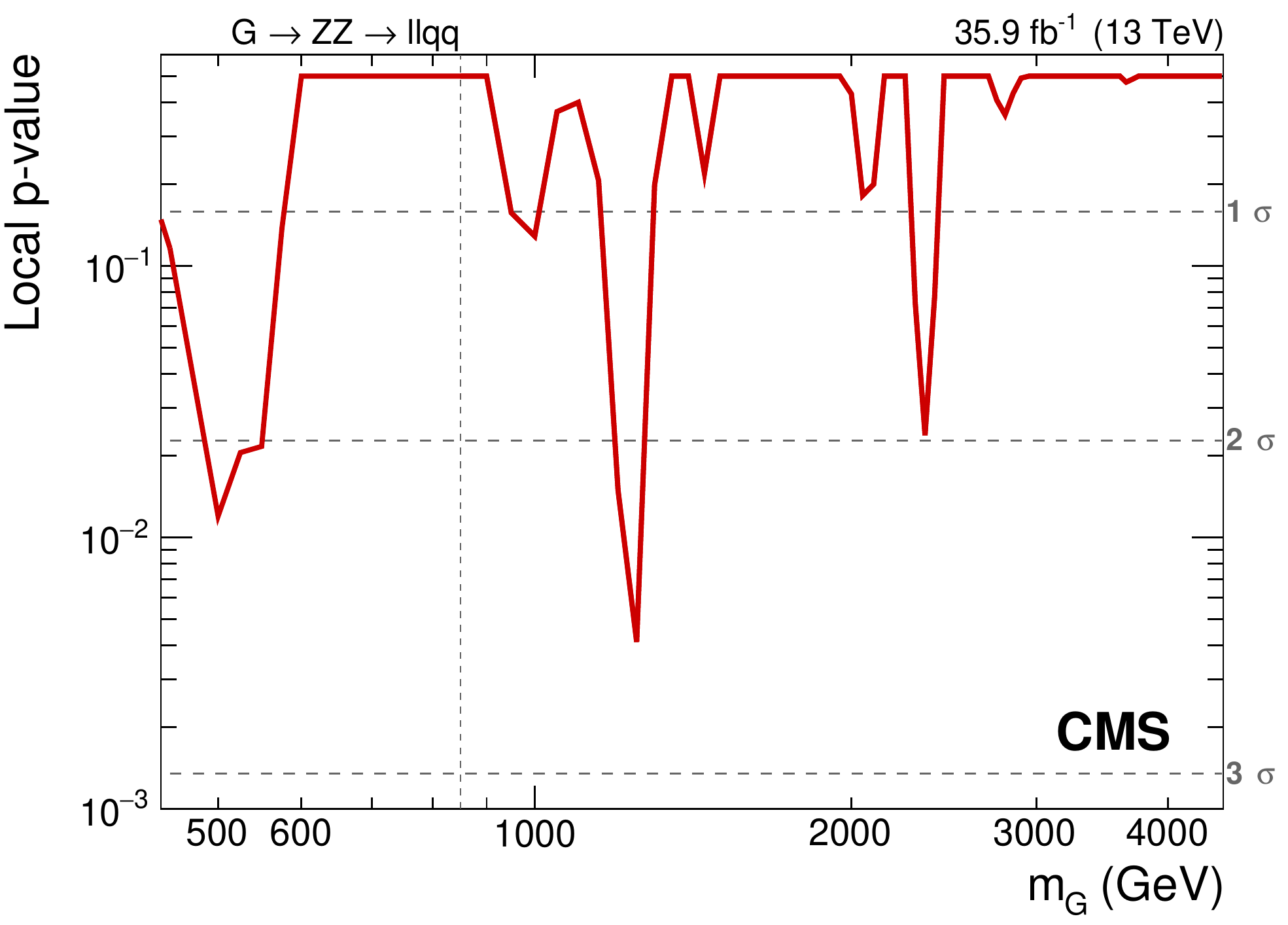}
    \caption{Observed local $p$-values for \PWpr (\cmsLeft) and {\cPG} (\cmsRight) narrow resonances as a function of the resonance mass. The dashed vertical line represents the transition from the low-mass to the high-mass analysis strategy.}
  \label{fig:pval}
\end{figure}

The observed upper limits on the resonance cross section, multiplied by the branching fraction for the decay into one \PZ boson and a {\PW} or \PZ boson, $\sigma_{\PWpr} \B(\PWpr\to\ZW)$ or $\sigma_{\cPG} \B(\cPG\to\ZZ)$, are reported as a function of the resonance mass in Fig.~\ref{fig:limit} assuming a \PWpr or {\cPG} produced in the narrow-width approximation, and the local $p$-value~\cite{pvalue} is shown in Fig.~\ref{fig:pval}.

Based on the observed (expected) upper limits on the signal cross section, a \PWpr signal is excluded up to 2270 (2390)\GeV in the framework of HVT model A ($g_{\mathrm{V}}=1$), and up to 2330 (2630)\GeV for HVT model B ($g_{\mathrm{V}}=3$); a WED bulk graviton is excluded up to masses of 925 (960)\GeV for $\ktilde=0.5$.

\section{Summary}\label{sec:conclusions}

A search for a heavy resonance decaying into a \PZ boson and a \PZ or a {\PW} boson in \llqq final states has been presented.
The data analyzed were collected by the CMS experiment in proton-proton collisions at $\sqrt{s}=13\TeV$ during 2016 operations at the LHC, corresponding to an integrated luminosity of 35.9\fbinv.
The final state of interest consists of a \PZ boson decaying leptonically into an electron or muon pair, and the decay of an additional {\PW} or \PZ boson into a pair of quarks.
Two analysis strategies, dedicated to the low- and high-mass regimes (below and above 850\GeV, respectively), have been used to set limits in the range of resonance mass from 400 to 4500\GeV.
Depending on the resonance mass, expected upper limits of 3--3000 and 1.5--400\unit{fb} have been set on the product of the cross section of a spin-1 \PWpr and the \ZW branching fraction, and on the product of the cross section of a spin-2 graviton and the \ZZ branching fraction, respectively.

\begin{acknowledgments}\label{sec:acknowledge}

We congratulate our colleagues in the CERN accelerator departments for the excellent performance of the LHC and thank the technical and administrative staffs at CERN and at other CMS institutes for their contributions to the success of the CMS effort. In addition, we gratefully acknowledge the computing centers and personnel of the Worldwide LHC Computing Grid for delivering so effectively the computing infrastructure essential to our analyses. Finally, we acknowledge the enduring support for the construction and operation of the LHC and the CMS detector provided by the following funding agencies: BMWFW and FWF (Austria); FNRS and FWO (Belgium); CNPq, CAPES, FAPERJ, and FAPESP (Brazil); MES (Bulgaria); CERN; CAS, MoST, and NSFC (China); COLCIENCIAS (Colombia); MSES and CSF (Croatia); RPF (Cyprus); SENESCYT (Ecuador); MoER, ERC IUT, and ERDF (Estonia); Academy of Finland, MEC, and HIP (Finland); CEA and CNRS/IN2P3 (France); BMBF, DFG, and HGF (Germany); GSRT (Greece); NKFIA (Hungary); DAE and DST (India); IPM (Iran); SFI (Ireland); INFN (Italy); MSIP and NRF (Republic of Korea); LAS (Lithuania); MOE and UM (Malaysia); BUAP, CINVESTAV, CONACYT, LNS, SEP, and UASLP-FAI (Mexico); MBIE (New Zealand); PAEC (Pakistan); MSHE and NSC (Poland); FCT (Portugal); JINR (Dubna); MON, RosAtom, RAS and RFBR (Russia); MESTD (Serbia); SEIDI, CPAN, PCTI and FEDER (Spain); Swiss Funding Agencies (Switzerland); MST (Taipei); ThEPCenter, IPST, STAR, and NSTDA (Thailand); TUBITAK and TAEK (Turkey); NASU and SFFR (Ukraine); STFC (United Kingdom); DOE and NSF (USA).

\hyphenation{Rachada-pisek} Individuals have received support from the Marie-Curie program and the European Research Council and Horizon 2020 Grant, contract No. 675440 (European Union); the Leventis Foundation; the A. P. Sloan Foundation; the Alexander von Humboldt Foundation; the Belgian Federal Science Policy Office; the Fonds pour la Formation \`a la Recherche dans l'Industrie et dans l'Agriculture (FRIA-Belgium); the Agentschap voor Innovatie door Wetenschap en Technologie (IWT-Belgium); the F.R.S.-FNRS and FWO (Belgium) under the ``Excellence of Science - EOS'' - be.h project n. 30820817; the Ministry of Education, Youth and Sports (MEYS) of the Czech Republic; the Lend\"ulet (``Momentum'') Program and the J\'anos Bolyai Research Scholarship of the Hungarian Academy of Sciences, the New National Excellence Program \'UNKP, the NKFIA research grants 123842, 123959, 124845, 124850 and 125105 (Hungary); the Council of Science and Industrial Research, India; the HOMING PLUS program of the Foundation for Polish Science, cofinanced from European Union, Regional Development Fund, the Mobility Plus program of the Ministry of Science and Higher Education, the National Science Center (Poland), contracts Harmonia 2014/14/M/ST2/00428, Opus 2014/13/B/ST2/02543, 2014/15/B/ST2/03998, and 2015/19/B/ST2/02861, Sonata-bis 2012/07/E/ST2/01406; the National Priorities Research Program by Qatar National Research Fund; the Programa Estatal de Fomento de la Investigaci{\'o}n Cient{\'i}fica y T{\'e}cnica de Excelencia Mar\'{\i}a de Maeztu, grant MDM-2015-0509 and the Programa Severo Ochoa del Principado de Asturias; the Thalis and Aristeia programs cofinanced by EU-ESF and the Greek NSRF; the Rachadapisek Sompot Fund for Postdoctoral Fellowship, Chulalongkorn University and the Chulalongkorn Academic into Its 2nd Century Project Advancement Project (Thailand); the Welch Foundation, contract C-1845; and the Weston Havens Foundation (USA).

\end{acknowledgments}

\bibliography{auto_generated}
\cleardoublepage \appendix\section{The CMS Collaboration \label{app:collab}}\begin{sloppypar}\hyphenpenalty=5000\widowpenalty=500\clubpenalty=5000\input{B2G-17-013-authorlist.tex}\end{sloppypar}
\end{document}

%% file: B2G-17-013-authorlist.tex
\vskip\cmsinstskip
\textbf{Yerevan~Physics~Institute, Yerevan, Armenia}\\*[0pt]
A.M.~Sirunyan, A.~Tumasyan
\vskip\cmsinstskip
\textbf{Institut~f\"{u}r~Hochenergiephysik, Wien, Austria}\\*[0pt]
W.~Adam, F.~Ambrogi, E.~Asilar, T.~Bergauer, J.~Brandstetter, E.~Brondolin, M.~Dragicevic, J.~Er\"{o}, A.~Escalante~Del~Valle, M.~Flechl, M.~Friedl, R.~Fr\"{u}hwirth\cmsAuthorMark{1}, V.M.~Ghete, J.~Grossmann, J.~Hrubec, M.~Jeitler\cmsAuthorMark{1}, A.~K\"{o}nig, N.~Krammer, I.~Kr\"{a}tschmer, D.~Liko, T.~Madlener, I.~Mikulec, E.~Pree, N.~Rad, H.~Rohringer, J.~Schieck\cmsAuthorMark{1}, R.~Sch\"{o}fbeck, M.~Spanring, D.~Spitzbart, A.~Taurok, W.~Waltenberger, J.~Wittmann, C.-E.~Wulz\cmsAuthorMark{1}, M.~Zarucki
% ...
\vskip\cmsinstskip
% ...
1:~Also at~Vienna~University~of~Technology, Vienna, Austria\\
% ...


%% file: B2G-17-013_temp.bbl
\providecommand{\href}[2]{#2}\begingroup\raggedright\begin{thebibliography}{10}%
\makeatletter
\providecommand{\hrefCMSnoop }[0]{\@secondoftwo}%
\makeatother
\providecommand{\doi}{\texttt{doi:}\begingroup \urlstyle{tt}\Url}

\bibitem{Agashe:2007zd}
\hrefCMSnoop {}{K.~Agashe, H.~Davoudiasl, G.~Perez, and A.~Soni, ``Warped
  gravitons at the {LHC} and beyond'',} \textit{ Phys. Rev. D} \textbf{ 76}
  (2007) 036006,
  \href{http://dx.doi.org/10.1103/PhysRevD.76.036006}{\doi{10.1103/PhysRevD.76.036006}},
\href{http://www.arXiv.org/abs/hep-ph/0701186}{\texttt{arXiv:hep-ph/0701186}}.

\bibitem{Randall:1999ee}
\hrefCMSnoop {}{L.~Randall and R.~Sundrum, ``A large mass hierarchy from a
  small extra dimension'',} \textit{ Phys. Rev. Lett.} \textbf{ 83} (1999)
  3370,
  \href{http://dx.doi.org/10.1103/PhysRevLett.83.3370}{\doi{10.1103/PhysRevLett.83.3370}},
\href{http://www.arXiv.org/abs/hep-ph/9905221}{\texttt{arXiv:hep-ph/9905221}}.

\bibitem{Randall:1999vf}
\hrefCMSnoop {}{L.~Randall and R.~Sundrum, ``An alternative to
  compactification'',} \textit{ Phys. Rev. Lett.} \textbf{ 83} (1999) 4690,
  \href{http://dx.doi.org/10.1103/PhysRevLett.83.4690}{\doi{10.1103/PhysRevLett.83.4690}},
\href{http://www.arXiv.org/abs/hep-th/9906064}{\texttt{arXiv:hep-th/9906064}}.

\bibitem{Fitzpatrick:2007qr}
\hrefCMSnoop {}{A.~L. Fitzpatrick, J.~Kaplan, L.~Randall, and L.-T. Wang,
  ``Searching for the {Kaluza-Klein} graviton in bulk {RS} models'',} \textit{
  JHEP} \textbf{ 09} (2007) 013,
  \href{http://dx.doi.org/10.1088/1126-6708/2007/09/013}{\doi{10.1088/1126-6708/2007/09/013}},
\href{http://www.arXiv.org/abs/hep-ph/0701150}{\texttt{arXiv:hep-ph/0701150}}.

\bibitem{Goldberger:1999uk}
\hrefCMSnoop {}{W.~D. Goldberger and M.~B. Wise, ``Modulus stabilization with
  bulk fields'',} \textit{ Phys. Rev. Lett.} \textbf{ 83} (1999) 4922,
  \href{http://dx.doi.org/10.1103/PhysRevLett.83.4922}{\doi{10.1103/PhysRevLett.83.4922}},
\href{http://www.arXiv.org/abs/hep-ph/9907447}{\texttt{arXiv:hep-ph/9907447}}.

\bibitem{Pappadopulo:2014qza}
\hrefCMSnoop {}{D.~Pappadopulo, A.~Thamm, R.~Torre, and A.~Wulzer, ``Heavy
  vector triplets: Bridging theory and data'',} \textit{ JHEP} \textbf{ 09}
  (2014) 060,
  \href{http://dx.doi.org/10.1007/JHEP09(2014)060}{\doi{10.1007/JHEP09(2014)060}},
\href{http://www.arXiv.org/abs/1402.4431}{\texttt{arXiv:1402.4431}}.

\bibitem{Khachatryan:2016cfx}
\hrefCMSnoop {}{{CMS Collaboration}, ``Search for heavy resonances decaying
  into a vector boson and a {Higgs} boson in final states with charged leptons,
  neutrinos, and b quarks'',} \textit{ Phys. Lett. B} \textbf{ 768} (2017) 137,
  \href{http://dx.doi.org/10.1016/j.physletb.2017.02.040}{\doi{10.1016/j.physletb.2017.02.040}},
\href{http://www.arXiv.org/abs/1610.08066}{\texttt{arXiv:1610.08066}}.

\bibitem{Sirunyan:2016cao}
\hrefCMSnoop {}{{CMS Collaboration}, ``Search for massive resonances decaying
  into {WW}, {WZ} or {ZZ} bosons in proton-proton collisions at {$\sqrt{s} = $
  13 TeV}'',} \textit{ JHEP} \textbf{ 03} (2017) 162,
  \href{http://dx.doi.org/10.1007/JHEP03(2017)162}{\doi{10.1007/JHEP03(2017)162}},
\href{http://www.arXiv.org/abs/1612.09159}{\texttt{arXiv:1612.09159}}.

\bibitem{Sirunyan:2017nrt}
\hrefCMSnoop {}{{CMS Collaboration}, ``Combination of searches for heavy
  resonances decaying to {WW}, {WZ}, {ZZ}, {WH}, and {ZH} boson pairs in
  proton-proton collisions at {$\sqrt{s}=8$} and 13 {TeV}'',} \textit{ Phys.
  Lett. B} \textbf{ 774} (2017) 533,
  \href{http://dx.doi.org/10.1016/j.physletb.2017.09.083}{\doi{10.1016/j.physletb.2017.09.083}},
\href{http://www.arXiv.org/abs/1705.09171}{\texttt{arXiv:1705.09171}}.

\bibitem{Aaboud:2017ahz}
\hrefCMSnoop {}{{ATLAS Collaboration}, ``Search for heavy resonances decaying
  to a {$W$} or {$Z$} boson and a {Higgs} boson in the
  $q\bar{q}^{(\prime)}b\bar{b}$ final state in $pp$ collisions at $\sqrt{s} =
  13$ {TeV} with the {ATLAS} detector'',} \textit{ Phys. Lett. B} \textbf{ 774}
  (2017) 494,
  \href{http://dx.doi.org/10.1016/j.physletb.2017.09.066}{\doi{10.1016/j.physletb.2017.09.066}},
\href{http://www.arXiv.org/abs/1707.06958}{\texttt{arXiv:1707.06958}}.

\bibitem{Sirunyan:2018ivv}
\hrefCMSnoop {}{{CMS Collaboration}, ``Search for a heavy resonance decaying
  into a {Z} boson and a vector boson in the
  $\nu\overline{\nu}\mathrm{q}\overline{\mathrm{q}}$ final state'',} \textit{
  JHEP} \textbf{ 07} (2018) 075,
  \href{http://dx.doi.org/10.1007/JHEP07(2018)075}{\doi{10.1007/JHEP07(2018)075}},
\href{http://www.arXiv.org/abs/1803.03838}{\texttt{arXiv:1803.03838}}.

\bibitem{MADGRAPH}
J.~Alwall\hrefCMSnoop {}{ {et~al.}, ``The automated computation of tree-level
  and next-to-leading order differential cross sections, and their matching to
  parton shower simulations'',} \textit{ JHEP} \textbf{ 07} (2014) 079,
  \href{http://dx.doi.org/10.1007/JHEP07(2014)079}{\doi{10.1007/JHEP07(2014)079}},
\href{http://www.arXiv.org/abs/1405.0301}{\texttt{arXiv:1405.0301}}.

\bibitem{Oliveira:2014kla}
\hrefCMSnoop {}{A.~Carvalho, ``Gravity particles from warped extra dimensions,
  predictions for {LHC}'',} (2014).
\href{http://www.arXiv.org/abs/1404.0102}{\texttt{arXiv:1404.0102}}.

\bibitem{FXFX}
\hrefCMSnoop {}{R.~Frederix and S.~Frixione, ``Merging meets matching in
  {MC@NLO}'',} \textit{ JHEP} \textbf{ 12} (2012) 061,
  \href{http://dx.doi.org/10.1007/JHEP12(2012)061}{\doi{10.1007/JHEP12(2012)061}},
\href{http://www.arXiv.org/abs/1209.6215}{\texttt{arXiv:1209.6215}}.

\bibitem{FEWZ}
\hrefCMSnoop {}{R.~Gavin, Y.~Li, F.~Petriello, and S.~Quackenbush, ``{FEWZ}
  2.0: A code for hadronic {Z} production at next-to-next-to-leading order'',}
  \textit{ Comput. Phys. Commun.} \textbf{ 182} (2011) 2388,
  \href{http://dx.doi.org/10.1016/j.cpc.2011.06.008}{\doi{10.1016/j.cpc.2011.06.008}},
\href{http://www.arXiv.org/abs/1011.3540}{\texttt{arXiv:1011.3540}}.

\bibitem{Alwall:2007fs}
J.~Alwall\hrefCMSnoop {}{ {et~al.}, ``Comparative study of various algorithms
  for the merging of parton showers and matrix elements in hadronic
  collisions'',} \textit{ Eur. Phys. J. C} \textbf{ 53} (2008) 473,
  \href{http://dx.doi.org/10.1140/epjc/s10052-007-0490-5}{\doi{10.1140/epjc/s10052-007-0490-5}},
\href{http://www.arXiv.org/abs/0706.2569}{\texttt{arXiv:0706.2569}}.

\bibitem{Nason:2004rx}
\hrefCMSnoop {}{P.~Nason, ``A new method for combining {NLO QCD} with shower
  {Monte Carlo} algorithms'',} \textit{ JHEP} \textbf{ 11} (2004) 040,
  \href{http://dx.doi.org/10.1088/1126-6708/2004/11/040}{\doi{10.1088/1126-6708/2004/11/040}},
\href{http://www.arXiv.org/abs/hep-ph/0409146}{\texttt{arXiv:hep-ph/0409146}}.

\bibitem{Frixione:2007vw}
\hrefCMSnoop {}{S.~Frixione, P.~Nason, and C.~Oleari, ``Matching {NLO} {QCD}
  computations with parton shower simulations: the {POWHEG} method'',} \textit{
  JHEP} \textbf{ 11} (2007) 070,
  \href{http://dx.doi.org/10.1088/1126-6708/2007/11/070}{\doi{10.1088/1126-6708/2007/11/070}},
\href{http://www.arXiv.org/abs/0709.2092}{\texttt{arXiv:0709.2092}}.

\bibitem{Alioli:2010xd}
\hrefCMSnoop {}{S.~Alioli, P.~Nason, C.~Oleari, and E.~Re, ``A general
  framework for implementing {NLO} calculations in shower {Monte Carlo}
  programs: the {POWHEG} {BOX}'',} \textit{ JHEP} \textbf{ 06} (2010) 043,
  \href{http://dx.doi.org/10.1007/JHEP06(2010)043}{\doi{10.1007/JHEP06(2010)043}},
\href{http://www.arXiv.org/abs/1002.2581}{\texttt{arXiv:1002.2581}}.

\bibitem{POWHEG_ST-st}
\hrefCMSnoop {}{S.~Alioli, P.~Nason, C.~Oleari, and E.~Re, ``{NLO} single-top
  production matched with shower in {POWHEG}: $s$- and $t$-channel
  contributions'',} \textit{ JHEP} \textbf{ 09} (2009) 111,
  \href{http://dx.doi.org/10.1088/1126-6708/2009/09/111}{\doi{10.1088/1126-6708/2009/09/111}},
  \href{http://www.arXiv.org/abs/0907.4076}{\texttt{arXiv:0907.4076}}.
[Erratum: \DOI{10.1007/JHEP02(2010)011}].

\bibitem{POWHEG_ST-tW}
\hrefCMSnoop {}{E.~Re, ``Single-top {Wt-channel} production matched with parton
  showers using the {POWHEG} method'',} \textit{ Eur. Phys. J. C} \textbf{ 71}
  (2011) 1547,
  \href{http://dx.doi.org/10.1140/epjc/s10052-011-1547-z}{\doi{10.1140/epjc/s10052-011-1547-z}},
\href{http://www.arXiv.org/abs/1009.2450}{\texttt{arXiv:1009.2450}}.

\bibitem{PYTHIA}
T.~Sj{\"o}strand\hrefCMSnoop {}{ {et~al.}, ``An introduction to {PYTHIA}
  8.2'',} \textit{ Comput. Phys. Commun.} \textbf{ 191} (2015) 159,
  \href{http://dx.doi.org/10.1016/j.cpc.2015.01.024}{\doi{10.1016/j.cpc.2015.01.024}},
\href{http://www.arXiv.org/abs/1410.3012}{\texttt{arXiv:1410.3012}}.

\bibitem{CUETP8M1}
\hrefCMSnoop {}{{CMS Collaboration}, ``Event generator tunes obtained from
  underlying event and multiparton scattering measurements'',} \textit{ Eur.
  Phys. J. C} \textbf{ 76} (2016) 155,
  \href{http://dx.doi.org/10.1140/epjc/s10052-016-3988-x}{\doi{10.1140/epjc/s10052-016-3988-x}},
\href{http://www.arXiv.org/abs/1512.00815}{\texttt{arXiv:1512.00815}}.

\bibitem{NNPDF}
\hrefCMSnoop {}{{NNPDF} Collaboration, ``Parton distributions from
  high-precision collider data'',} \textit{ Eur. Phys. J. C} \textbf{ 77}
  (2017) 663,
  \href{http://dx.doi.org/10.1140/epjc/s10052-017-5199-5}{\doi{10.1140/epjc/s10052-017-5199-5}},
\href{http://www.arXiv.org/abs/1706.00428}{\texttt{arXiv:1706.00428}}.

\bibitem{GEANT4}
\hrefCMSnoop {}{{GEANT4} Collaboration, ``{GEANT4}: A simulation toolkit'',}
  \textit{ Nucl. Instrum. Meth. A} \textbf{ 506} (2003) 250,
\href{http://dx.doi.org/10.1016/S0168-9002(03)01368-8}{\doi{10.1016/S0168-9002(03)01368-8}}.

\bibitem{Chatrchyan:2008zzk}
\hrefCMSnoop {}{{CMS Collaboration}, ``The {CMS} experiment at the {CERN}
  {LHC}'',} \textit{ JINST} \textbf{ 3} (2008) S08004,
\href{http://dx.doi.org/10.1088/1748-0221/3/08/S08004}{\doi{10.1088/1748-0221/3/08/S08004}}.

\bibitem{Sirunyan:2017ulk}
\hrefCMSnoop {}{{CMS Collaboration}, ``Particle-flow reconstruction and global
  event description with the {CMS} detector'',} \textit{ JINST} \textbf{ 12}
  (2017) P10003,
  \href{http://dx.doi.org/10.1088/1748-0221/12/10/P10003}{\doi{10.1088/1748-0221/12/10/P10003}},
\href{http://www.arXiv.org/abs/1706.04965}{\texttt{arXiv:1706.04965}}.

\bibitem{TRK-11-001}
\hrefCMSnoop {}{{CMS Collaboration}, ``Description and performance of track and
  primary-vertex reconstruction with the {CMS} tracker'',} \textit{ JINST}
  \textbf{ 9} (2014) P10009,
  \href{http://dx.doi.org/10.1088/1748-0221/9/10/P10009}{\doi{10.1088/1748-0221/9/10/P10009}},
\href{http://www.arXiv.org/abs/1405.6569}{\texttt{arXiv:1405.6569}}.

\bibitem{Khachatryan:2015hwa}
\hrefCMSnoop {}{{CMS Collaboration}, ``Performance of electron reconstruction
  and selection with the {CMS} detector in proton-proton collisions at
  {$\sqrt{s} = 8$\TeV}'',} \textit{ JINST} \textbf{ 10} (2015) P06005,
  \href{http://dx.doi.org/10.1088/1748-0221/10/06/P06005}{\doi{10.1088/1748-0221/10/06/P06005}},
\href{http://www.arXiv.org/abs/1502.02701}{\texttt{arXiv:1502.02701}}.

\bibitem{Chatrchyan:2012xi}
\hrefCMSnoop {}{{CMS Collaboration}, ``Performance of {CMS} muon reconstruction
  in pp collision events at {$\sqrt{s} = 7$\TeV}'',} \textit{ JINST} \textbf{
  7} (2012) P10002,
  \href{http://dx.doi.org/10.1088/1748-0221/7/10/P10002}{\doi{10.1088/1748-0221/7/10/P10002}},
\href{http://www.arXiv.org/abs/1206.4071}{\texttt{arXiv:1206.4071}}.

\bibitem{Cacciari:2008gp}
\hrefCMSnoop {}{M.~Cacciari, G.~P. Salam, and G.~Soyez, ``The anti-\kt jet
  clustering algorithm'',} \textit{ JHEP} \textbf{ 04} (2008) 063,
  \href{http://dx.doi.org/10.1088/1126-6708/2008/04/063}{\doi{10.1088/1126-6708/2008/04/063}},
  \href{http://www.arXiv.org/abs/0802.1189}{\texttt{arXiv:0802.1189}}.

\bibitem{Cacciari:2011ma}
\hrefCMSnoop {}{M.~Cacciari, G.~P. Salam, and G.~Soyez, ``{FastJet} user
  manual'',} \textit{ Eur. Phys. J. C} \textbf{ 72} (2012) 1896,
  \href{http://dx.doi.org/10.1140/epjc/s10052-012-1896-2}{\doi{10.1140/epjc/s10052-012-1896-2}},
\href{http://www.arXiv.org/abs/1111.6097}{\texttt{arXiv:1111.6097}}.

\bibitem{Khachatryan:2016kdb}
\hrefCMSnoop {}{{CMS Collaboration}, ``Jet energy scale and resolution in the
  {CMS} experiment in pp collisions at 8 {TeV}'',} \textit{ JINST} \textbf{ 12}
  (2017) P02014,
  \href{http://dx.doi.org/10.1088/1748-0221/12/02/P02014}{\doi{10.1088/1748-0221/12/02/P02014}},
\href{http://www.arXiv.org/abs/1607.03663}{\texttt{arXiv:1607.03663}}.

\bibitem{Dasgupta:2013ihk}
\hrefCMSnoop {}{M.~Dasgupta, A.~Fregoso, S.~Marzani, and G.~P. Salam, ``Towards
  an understanding of jet substructure'',} \textit{ JHEP} \textbf{ 09} (2013)
  029,
  \href{http://dx.doi.org/10.1007/JHEP09(2013)029}{\doi{10.1007/JHEP09(2013)029}},
\href{http://www.arXiv.org/abs/1307.0007}{\texttt{arXiv:1307.0007}}.

\bibitem{Larkoski:2014wba}
\hrefCMSnoop {}{A.~J. Larkoski, S.~Marzani, G.~Soyez, and J.~Thaler, ``Soft
  drop'',} \textit{ JHEP} \textbf{ 05} (2014) 146,
  \href{http://dx.doi.org/10.1007/JHEP05(2014)146}{\doi{10.1007/JHEP05(2014)146}},
\href{http://www.arXiv.org/abs/1402.2657}{\texttt{arXiv:1402.2657}}.

\bibitem{Bertolini2014}
\hrefCMSnoop {}{D.~Bertolini, P.~Harris, M.~Low, and N.~Tran, ``Pileup per
  particle identification'',} \textit{ JHEP} \textbf{ 10} (2014) 059,
  \href{http://dx.doi.org/10.1007/JHEP10(2014)059}{\doi{10.1007/JHEP10(2014)059}},
\href{http://www.arXiv.org/abs/1407.6013}{\texttt{arXiv:1407.6013}}.

\bibitem{Thaler2011}
\hrefCMSnoop {}{J.~Thaler and K.~Van~Tilburg, ``Identifying boosted objects
  with {N-subjettiness}'',} \textit{ JHEP} \textbf{ 03} (2011) 015,
  \href{http://dx.doi.org/10.1007/JHEP03(2011)015}{\doi{10.1007/JHEP03(2011)015}},
\href{http://www.arXiv.org/abs/1011.2268}{\texttt{arXiv:1011.2268}}.

\bibitem{btag}
\hrefCMSnoop {}{{CMS Collaboration}, ``Identification of b-quark jets with the
  {CMS} experiment'',} \textit{ JINST} \textbf{ 8} (2013) P04013,
  \href{http://dx.doi.org/10.1088/1748-0221/8/04/P04013}{\doi{10.1088/1748-0221/8/04/P04013}},
\href{http://www.arXiv.org/abs/1211.4462}{\texttt{arXiv:1211.4462}}.

\bibitem{CMS-PAS-BTV-15-001}
\href {https://cds.cern.ch/record/2138504}{{CMS Collaboration},
  ``Identification of b quark jets at the {CMS} experiment in the {LHC} {Run
  2}'',} CMS Physics Analysis Summary CMS-PAS-BTV-15-001, 2016.

\bibitem{Khachatryan:2016bia}
\hrefCMSnoop {}{{CMS Collaboration}, ``{The CMS trigger system}'',} \textit{
  JINST} \textbf{ 12} (2017) P01020,
  \href{http://dx.doi.org/10.1088/1748-0221/12/01/P01020}{\doi{10.1088/1748-0221/12/01/P01020}},
\href{http://www.arXiv.org/abs/1609.02366}{\texttt{arXiv:1609.02366}}.

\bibitem{Sirunyan:2018fpa}
\hrefCMSnoop {}{{CMS Collaboration}, ``Performance of the {CMS} muon detector
  and muon reconstruction with proton-proton collisions at $\sqrt{s}=$ 13
  {TeV}'',} \textit{ JINST} \textbf{ 13} (2018) P06015,
  \href{http://dx.doi.org/10.1088/1748-0221/13/06/P06015}{\doi{10.1088/1748-0221/13/06/P06015}},
\href{http://www.arXiv.org/abs/1804.04528}{\texttt{arXiv:1804.04528}}.

\bibitem{Sirunyan:2018qob}
\hrefCMSnoop {}{{CMS Collaboration}, ``Search for heavy resonances decaying
  into a vector boson and a {Higgs} boson in final states with charged leptons,
  neutrinos and b quarks at $\sqrt{s}=$ 13 {TeV}'',} (2018).
  \href{http://www.arXiv.org/abs/1807.02826}{\texttt{arXiv:1807.02826}}.
Submitted to {\it JHEP}.

\bibitem{CBfunction}
\href {https://www.slac.stanford.edu/cgi-wrap/getdoc/slac-r-236.pdf}{M.~J.
  Oreglia, ``A study of the reactions $\psi^\prime \to \gamma \gamma \psi$''}.
\newblock PhD thesis, Stanford University, 1980.
\newblock
SLAC Report SLAC-236.

\bibitem{Khachatryan:2016txa}
\hrefCMSnoop {}{{CMS Collaboration}, ``Measurement of the {ZZ} production cross
  section and {Z~$\to \ell^+\ell^-\ell'^+\ell'^-$} branching fraction in pp
  collisions at $\sqrt s$ = 13 {TeV}'',} \textit{ Phys. Lett. B} \textbf{ 763}
  (2016) 280,
  \href{http://dx.doi.org/10.1016/j.physletb.2016.10.054}{\doi{10.1016/j.physletb.2016.10.054}},
\href{http://www.arXiv.org/abs/1607.08834}{\texttt{arXiv:1607.08834}}.

\bibitem{Khachatryan:2016tgp}
\hrefCMSnoop {}{{CMS Collaboration}, ``Measurement of the {WZ} production cross
  section in pp collisions at $\sqrt{s} =$ 13 {TeV}'',} \textit{ Phys. Lett. B}
  \textbf{ 766} (2017) 268,
  \href{http://dx.doi.org/10.1016/j.physletb.2017.01.011}{\doi{10.1016/j.physletb.2017.01.011}},
\href{http://www.arXiv.org/abs/1607.06943}{\texttt{arXiv:1607.06943}}.

\bibitem{Khachatryan:2016kzg}
\hrefCMSnoop {}{{CMS Collaboration}, ``Measurement of the $\mathrm{t\bar{t}}$
  production cross section using events in the e$\mu$ final state in pp
  collisions at $\sqrt{s} =$ 13 {TeV}'',} \textit{ Eur. Phys. J. C} \textbf{
  77} (2017) 172,
  \href{http://dx.doi.org/10.1140/epjc/s10052-017-4718-8}{\doi{10.1140/epjc/s10052-017-4718-8}},
\href{http://www.arXiv.org/abs/1611.04040}{\texttt{arXiv:1611.04040}}.

\bibitem{CMS-PAS-JME-16-003}
\href {http://cds.cern.ch/record/2256875}{{CMS Collaboration}, ``Jet algorithms
  performance in 13 {TeV} data'',} CMS Physics Analysis Summary
  CMS-PAS-JME-16-003, 2017.

\bibitem{Khachatryan2014}
\hrefCMSnoop {}{{CMS Collaboration}, ``Identification techniques for highly
  boosted {W} bosons that decay into hadrons'',} \textit{ JHEP} \textbf{ 12}
  (2014) 017,
  \href{http://dx.doi.org/10.1007/JHEP12(2014)017}{\doi{10.1007/JHEP12(2014)017}},
\href{http://www.arXiv.org/abs/1410.4227}{\texttt{arXiv:1410.4227}}.

\bibitem{HERWIG}
M.~Bahr\hrefCMSnoop {}{ {et~al.}, ``Herwig++ physics and manual'',} \textit{
  Eur. Phys. J. C} \textbf{ 58} (2008) 639,
  \href{http://dx.doi.org/10.1140/epjc/s10052-008-0798-9}{\doi{10.1140/epjc/s10052-008-0798-9}},
\href{http://www.arXiv.org/abs/0803.0883}{\texttt{arXiv:0803.0883}}.

\bibitem{CMS-PAS-LUM-17-001}
\href {http://cds.cern.ch/record/2257069}{{CMS Collaboration}, ``{CMS}
  luminosity measurements for the 2016 data taking period'',} CMS Physics
  Analysis Summary CMS-PAS-LUM-17-001, 2017.

\bibitem{Read:2002hq}
\hrefCMSnoop {}{A.~L. Read, ``Presentation of search results: The
  {CL$_{\text{s}}$} technique'',} \textit{ J. Phys. G} \textbf{ 28} (2002)
  2693,
\href{http://dx.doi.org/10.1088/0954-3899/28/10/313}{\doi{10.1088/0954-3899/28/10/313}}.

\bibitem{Junk:1999kv}
\hrefCMSnoop {}{T.~Junk, ``Confidence level computation for combining searches
  with small statistics'',} \textit{ Nucl. Instrum. Meth. A} \textbf{ 434}
  (1999) 435,
  \href{http://dx.doi.org/10.1016/S0168-9002(99)00498-2}{\doi{10.1016/S0168-9002(99)00498-2}},
\href{http://www.arXiv.org/abs/hep-ex/9902006}{\texttt{arXiv:hep-ex/9902006}}.

\bibitem{Cowan:2010js}
\hrefCMSnoop {}{G.~Cowan, K.~Cranmer, E.~Gross, and O.~Vitells, ``Asymptotic
  formulae for likelihood-based tests of new physics'',} \textit{ Eur. Phys. J.
  C} \textbf{ 71} (2011) 1554,
  \href{http://dx.doi.org/10.1140/epjc/s10052-011-1554-0}{\doi{10.1140/epjc/s10052-011-1554-0}},
  \href{http://www.arXiv.org/abs/1007.1727}{\texttt{arXiv:1007.1727}}.
[Erratum: \DOI{10.1140/epjc/s10052-013-2501-z}].

\bibitem{CMS-NOTE-2011-005}
\href {https://cds.cern.ch/record/1379837}{{The ATLAS Collaboration, The CMS
  Collaboration, The LHC Higgs Combination Group}, ``Procedure for the {LHC}
  {Higgs} boson search combination in {Summer} 2011'',} Technical Report
  CMS-NOTE-2011-005. ATL-PHYS-PUB-2011-11, 2011.

\bibitem{pvalue}
\hrefCMSnoop {}{L.~Demortier, ``{P} values and nuisance parameters'',} in
  \textit{ Statistical issues for {LHC} physics. {Proceedings, Workshop,
  PHYSTAT-LHC, Geneva, Switzerland, June} 27-29, 2007}, p.~23.
\newblock 2008.
\newblock
\href{http://dx.doi.org/10.5170/CERN-2008-001}{\doi{10.5170/CERN-2008-001}}.

\end{thebibliography}\endgroup
